\documentclass[%
 reprint,
nofootinbib,
 amsmath,amssymb,
 aps,
]{revtex4-2}

\usepackage{graphicx}
\usepackage{dcolumn}
\usepackage{bm}

\usepackage{float}
\usepackage{mathptmx}
\usepackage{amsmath,amssymb}
\usepackage{mathtools}
\usepackage[T1]{fontenc}
\usepackage[utf8]{inputenc}
\usepackage[english]{babel}
\addto{\captionsenglish}{%
  
}
\usepackage{array}
\usepackage{droidsans}
\usepackage{charter}
\usepackage[usenames,dvipsnames]{xcolor}
\usepackage{setspace}
\usepackage[compact]{titlesec}
\usepackage{hyperref}

\begin{document}

\preprint{ }

\title{The Cellular Potts Model on Disordered Lattices}

\author{Hossein Nemati}
\email{h.nemati@uu.nl}
\author{Joost de Graaf}%
\affiliation{%
 Institute for Theoretical Physics, Center for Extreme Matter and Emergent Phenomena, Utrecht University, Princetonplein 5, 3584 CC Utrecht, The Netherlands.
}%
\date{\today}

\begin{abstract}
The Cellular Potts model, also known as the Glazier-Graner-Hogeweg model, is a lattice-based approach by which biological tissues at the level of individual cells can be numerically studied. Traditionally, a square or hexagonal underlying lattice structure is assumed for two-dimensional systems, and this is known to introduce artifacts in the structure and dynamics of the model tissues. That is, on regular lattices, cells can assume shapes that are dictated by the symmetries of the underlying lattice. Here, we developed a variant of this method that can be applied to a broad class of (ir)regular lattices. We show that on an irregular lattice deriving from a fluid-like configuration, two types of artifacts can be removed. We further report on the transition between a fluid-like disordered and a solid-like hexagonally ordered phase present for monodisperse confluent cells as a function of their surface tension. This transition shows the hallmarks of a first-order phase transition and is different from the glass/jamming transitions commonly reported for the vertex and active Voronoi models. We emphasize this by analyzing the distribution of shape parameters found in our state space. Our analysis provides a useful reference for the future study of epithelia using the (ir)regular Cellular Potts model.
\end{abstract}

\maketitle


\section{\label{sec:intro}Introduction}

Collective cell migration plays an essential role in many biological settings, ranging from morphogenesis~\cite{merkel2017, lin2018,  akiyama2017, Friedl2009, rorth2009}, to wound healing~\cite{de_leon2023, xu2023, brugues2014, poujade2007}, to cancer metastasis~\cite{Friedl2009, cheung2013, metzcar2019, Ilina2020, Grosser2021, Oswald2017, Blauth2021}. Confluent cell monolayers, such as those found in epithelial tissues, have attracted considerable attention from the modeling community. Firstly, their effective two-dimensional (2D) nature makes them relatively simple to study, yet these tissues can exhibit a wide variety of collective~\cite{poujade2007, cheung2013, yin2024, armengol2023} and rheological~\cite{Bi2015, Park2015, Damavandi2022, hopkins2022, fielding2023, Hertaeg2024} behaviors. Secondly, the fact that epithelia form the outer surfaces of our organs makes them susceptible to cancer~\cite{chamoli2021, levra2024} and play a role in a wide variety of diseases~\cite{okamoto2016, holtzman2014, klettner2020}, including pulmonary fibrosis~\cite{stancil2021} and asthma~\cite{Park2015}. This gives the study of epithelia a direct biomedical relevance.

Over the past decades, various quantitative (computational) models have been developed that capture the principal features of cell populations, without being overburdened by complexity~\cite{Alert2020, camley2017}. Many are agent-based in nature and have seen concurrent use in the study of active matter~\cite{shaebani2020}. For tissues and cells, these models include lattice-based methods like cellular Potts model (CPM)~\cite{Graner1992, Scianna2013, Scianna2012, Mare, Szab2013, Hirashima2017, Boas2018, HOGEWEG2000} and cellular automata (CA)~\cite{Poorkhanalikoudehi2021,ermentrout1993,hatzikirou2008, nava2017}, particle-based~\cite{kirchner2024, akiyama2017, camley2016, basan2013, campo2019}, vertex~\cite{Farhadifar2007, Bi2015, alt2017, fletcher2014, landsberg2009}, Voronoi~\cite{Bi2016, Giavazzi2018, Yang2017, pinto2022}, phase-field~\cite{moure2021, zhang2023, nonomura2012, loewe2020, jain2023, wenzel2021, Monfared2023}, and Fourier-contour models~\cite{saito2023active}. 
We should emphasize that the constraint of confluency is an important difference between particle-based models and models that have this feature built into their description, \textit{e.g.,} the vertex, Voronoi, and CPM.
Evolutionary dynamics can also be included in models to account for (cancerous) mutations and invasion~\cite{tkadlec2023, nemati2023, renton2021, renton2019, manem2015}. We refer to Refs.~\citenum{metzcar2019, Alert2020, moure2021, goriely2017} for overviews of the different models available and their respective ranges of application. Reference~\citenum{metzcar2019} provides a detailed list of open-source implementations of these models, should one wish to experiment. Lastly, Refs.~\citenum{beatrici2023, osborne2017} have made comparisons between different models, which included their ability to describe cell shapes and the way cells exchange neighbors.

In this paper, we will limit ourselves to the CPM, which has good performance on these two points, \textit{i.e.}, cell-shape descriptiveness, and natural neighbor exchange. These abilities of the CPM, naturally lead to a greater computational cost when compared to the models in which cells are described as single particles~\cite{beatrici2023, osborne2017}. However, the method is generally considered efficient and has seen widespread adoption. For example, the CPM has been successfully used to study cell sorting and rearrangements~\cite{Glazier1993, Graner1992, Mombach1995}, chemotaxis~\cite{tan2024, ouaknin2009, kafer2006}, topotaxis~\cite{steijn2023}, 
cell migration patterns~\cite{plazen2023, wortel2021, niculescu2015, matsushita2022, kabla2012}, wound closure~\cite{noppe2015, scianna2015}, tumor growth and invasion~\cite{Szab2013}, and the interactions of cells with extracellular matrix~\cite{rubenstein2008, tsingos2023, crossley2024}. Several open-source packages are available for using the CPM~\cite{swat2012, chaste_2020, simmune_1999, toolkit2015, morpheus2014, artistoo2021} and the approach has been extensively reviewed,~\textit{e.g.}, see Refs.~\citenum{savill2007, Mare, Scianna2012, Scianna2013}. However, despite its popularity and qualities, the method is known to suffer from lattice artifacts~\cite{Mare}. That is, in certain regimes of model parameters, the shape of cells and macroscopic properties of the system are strongly affected by the symmetries of the underlying lattice. 
Some of the known issues have been addressed. For example, cells can be fragmented at a sufficiently high rate of cell membrane fluctuations (low surface tension). This was resolved by Durand and Guesnet~\cite{Durand2016} by adding an efficient connectivity check. In addition, a node-based version of CPM has been recently proposed with the goal of reducing lattice artifacts~\cite{Scianna2016}. This model describes the cells as polygons and tracks their vertices.

Here, we will take a different route toward removing/reducing lattice artifacts. We employ an irregular and on-average homogeneous lattice to support our CPM, which derives from a separate simulation of a fluid-like state \textit{via} Voronoi tessellation. Using this supporting lattice, we study how the absence of long-range order in the lattice affects the transition between (disordered) fluid-like and solid-like states that can occur in model epithelia. Such changes of state are experimentally reported to include\footnote{ It should be mentioned that, in the context of condensed matter, the \textit{glass} and \textit{jamming} transition are different and have distinct underlying physics. However, here, we chose to include all the nomenclature that is used in the literature of tissue mechanics without judging the accuracy of the specific use.} \textit{jamming}, \textit{rigidity}, and \textit{glass transitions}~\cite{Oswald2017, nnetu2012, schotz2013, bocanegra-moreno2023, Park2015, Hannezo2022, Garcia2015, LawsonKeister2021, angelini2011}, and reproduced in a variety of vertex- and Voronoi-based computational studies~\cite{Bi2015, Bi2016, bi2014, Thomas2023, pandey2023}, which also include polydisperse systems\cite{sussman2018}. 
Here, we should mention that in Voronoi-based models, although dynamical transitions are observed\cite{Bi2016}, they are known to be absent in the athermal version of the model\cite{sussman_no_2018}. 
These transitions have attracted attention as key players in the development of the aforementioned tissue-related diseases. The change from fluid-like to solid-like is commonly referred to as a \textit{phase transition} in the literature~\cite{lenne2022, petridou2021, Oswald2017, Park2015, LawsonKeister2021} and we adopt the nomenclature. Note, however, that biological tissues are intrinsically out of equilibrium.

Disordered arrested dynamics and fluid-to-solid transitions have been reported for the CPM~\cite{Chiang2016, Sadhukhan2021, Devanny2023}. However, Durand and Heu~\cite{Durand2019} used the CPM to study soft cellular systems and instead found an order-to-disorder phase transition. We revisit the work by Chiang and Marenduzzo~\cite{Chiang2016} and show that lattice artifacts present in their systems are removed by our irregular-lattice CPM. Our results further demonstrate that for the CPM with their (simple) Hamiltonian, there is an order-to-disorder transition rather than a disordered solidification. This transition has all the hallmarks of a first-order phase transition from a fluid to a hexagonal solid. We verified the nature of the transition by examining in detail the geometric features of the cells in the tissue and their neighborhoods. Such features include the (distribution of) the isoperimetric quotient and the circularity. We have also studied how our irregular lattice compares to the use of a hexagonal one and find that the latter has spurious dynamics in the hexagonal crystal state.

The benefits of using a CPM on an irregular lattice come at only a small computational overhead compared to using the regular CPM. This makes it a suitable alternative for the study of real biological tissues, which we aim to pursue in future work.

%

\section{\label{sec:method}Methodology}

In this section, we cover the main features of our variant of CPM: the creation and characterization of irregular lattices, and the means by which we have modified the traditional CPM to work with these lattices. We also discuss the various means, including the mean-squared displacement (MSD) and isoperimetric quotient, by which we characterize the outcomes of our simulations. We further provide our standard choices for the system parameters and simulation ranges that we have considered.

\subsection{\label{sub:CPM}The Cellular Potts Model}

We introduce the basic CPM algorithm here so that the background for our extension is set. Assume that we have a lattice with $N_{s}$ sites and $N_{c}$ cells, we can define a function $\sigma :\{1,\dots,N_{s}\} \xmapsto{}\{1,\dots,N_{c}\}$ that describes the configuration of the cells on the lattice, uniquely. That is, $\sigma(i)$ indicates the cell index, to which the site $i$ of the lattice belongs. As the development of CPM was inspired by the Potts model~\cite{Potts1952} --- originally used to study spin systems --- let us call $\sigma(i)$ the \textit{spin} of site $i$. Note that the real biological system of interest for the CPM has nothing to do with magnetism and the spin indices.

Now that we can describe a configuration using $\sigma$, we can specify the Hamiltonian that gives the total energy of the system, $H= H(\sigma, \mathcal{P})$, where $\mathcal{P}$ represents the set of all physical parameters of the system,~\textit{e.g.}, surface tension and target cell area. In general, the lattice can be of any dimension, but we will restrict ourselves to 2D CPMs here. These are suited to describe cell monolayers, which form a class of epithelia, found in the cells lining blood vessels and alveolar sacs of the lung~\cite{kuhnel_102_2003}. The simplest form of Hamiltonian that is used on 2D regular lattices is given by
\begin{align}
\label{eq:hamiltonian_reg} H &= \frac{\alpha}{2} \sum^{N_{s}}_{i=1} \sum_{j \in \mathcal{N}(i)} \left( 1 - \delta_{\sigma(i),\sigma(j)} \right) + \lambda\sum_{\sigma=1}^{N_{c}} \left( a_{\sigma} - A_{0} \right)^{2} .
\end{align}

The first term gives the total interaction energy between the cells that comes from the surface tension between the cell membranes. The indices $i$ and $j$ indicate the lattice sites and the summation is carried out over the site pairs within each other's \textit{interaction neighborhood}. The Kronecker delta $\delta_{ij}$ ($1$ if $i = j$ and $0$ if $i \ne j$ for any indices) is used to indicate that only adjacent sites that belong to cells with different spins contribute to the surface tension. The factor $\alpha$ indicates the surface tension and is typically considered uniform between cells, though the method can be used to study mixtures of different cell types as well~\cite{Glazier1993, Graner1992, Steinberg2007}.

The second term indicates that each cell has a preferred (or target) area $A_{0}$. Departures of the instantaneous area $a_{\sigma}$ away from $A_{0}$ are penalized using a Hookean potential with spring constant $2\lambda$. This choice models the tendency of cells to have a constant volume, as well as their connection to their neighbors within an epithelium. That is, any change in shape (elongation / shrinking out of the plane) will lead to a change in the in-plane area (volume conservation). It is assumed that any deviation from the natural shape is associated with an energy cost, which at the lowest order would be quadratic.

The `dynamics' of the CPM is propagated using a Monte-Carlo (MC) approach. This consists of trial moves weighted by the change in the energy. A trial move involves changing the spin of a randomly chosen site to that of its neighbors. In brief, the basic implementation is
\begin{enumerate}
    \item Store the energy of the current configuration $E_{\mathrm{old}} = H(\sigma_{\mathrm{old}}, \mathcal{P})$ and choose a lattice site randomly, say, site $i$. We call it the \textit{candidate site}.
    \item Choose a site from its surroundings\footnote{This neighborhood $\mathcal{N}_{c}(i)$ is not necessarily the same as the one used for the computation of the Hamiltonian $\mathcal{N}(i)$ and may be defined separately.}, $\mathcal{N}_{c}(i)$, say, site $j$. We call it the \textit{invading site}.
    \item If $\sigma(i) = \sigma(j)$, go to step 1. Otherwise, temporarily change $\sigma(i)$ into $\sigma(j)$, and calculate the new value of energy, $E_{\mathrm{new}} = H(\sigma_{\mathrm{new}}, \mathcal{P})$. We call this configurational change an \textit{attempt}.
    \item Call the change in energy $\Delta E = E_{\mathrm{new}} - E_{\mathrm{old}}$. When $\Delta E \leq 0$, accept the attempt. When $\Delta E > 0$, accept the attempt with the probability
    \begin{align}
    \label{eq:accept} P_{\mathrm{acc}} &= \exp \left(- \frac{ \Delta E }{ k_{\mathrm{B}}T } \right) ,
    \end{align}
    or reject it with probability $1 - P_{\mathrm{acc}}$. Here, the energy difference is normalized by the thermal energy, where $k_{\mathrm{B}}$ is the Boltzmann constant and $T$ is the temperature. These steps will repeat until a user-defined maximum number of iterations is reached.
\end{enumerate}
It should be noted that the use of thermal energy in the algorithm, reflects its origin as a tool to study spin systems. The interpretation of the temperature in the context of a cell membrane is to set a rate, at which the various (internal) cell activities change the boundaries. This rate can be given the interpretation of a time scale for membrane fluctuations~\cite{Mombach1995, Durand2016, Mare}, provided only local trial moves are used. Nonetheless, the CPM's dynamics do not represent the evolution of the system, as would follow from say a Langevin description. 
However, we should note that there have been attempts to reconcile the `time' in the CPM with a physically meaningful time through the introduction of Poissonian statistics~\cite{belousov2024}.

According to the algorithm that we have described above, the site pairs which belong to the same cell, are also allowed to be chosen. Such picks are always disregarded for updates after checking that the spins are the same. However, picking these pairs comes at a computational cost. It is possible to identify all `allowed site pairs' in \textit{linked list} data structures~\cite{Antonakos1999} and only choose from the elements of these lists. However, in practice, for our typical parameter choices, it turned out that searching and updating the linked lists was computationally disadvantageous. Hence, we utilized the above algorithm. Note that linked lists should become more efficient when the number of border sites is considerably less than the total number of lattice sites. Examples of this include the simulation of single cells~\cite{allena2016} and non-confluent cell populations~\cite{guisoni2018, rens2019}.

\subsection{\label{sub:lattice} Using (Ir)regular Lattices}

The above algorithm for site updates holds for \textit{any} underlying lattice, provided $H(\sigma, \mathcal{P})$ and $\mathcal{N}_c(i)$ for $1\leq i \leq N_{s}$ are well-defined. As mentioned in the introduction, CPM has predominantly been applied to regular lattices,~\textit{i.e.}, square and hexagonal lattices. We are aware of only one instance of a study of an irregular lattice mentioned in the literature\cite{leoncini2010}; in Ref.~\citenum{vanliedekerke2015} a graph is shown of what appears to be a study performed on an irregular lattice, but the original text could not be obtained. For the sake of completeness, we should also mention the node-based version of CPM~\cite{Scianna2016}, which describes cells through surfaces rather than through volumes, making it principally different from the standard CPM.

\begin{figure}[!htb]
\centering
\includegraphics[width=\columnwidth]{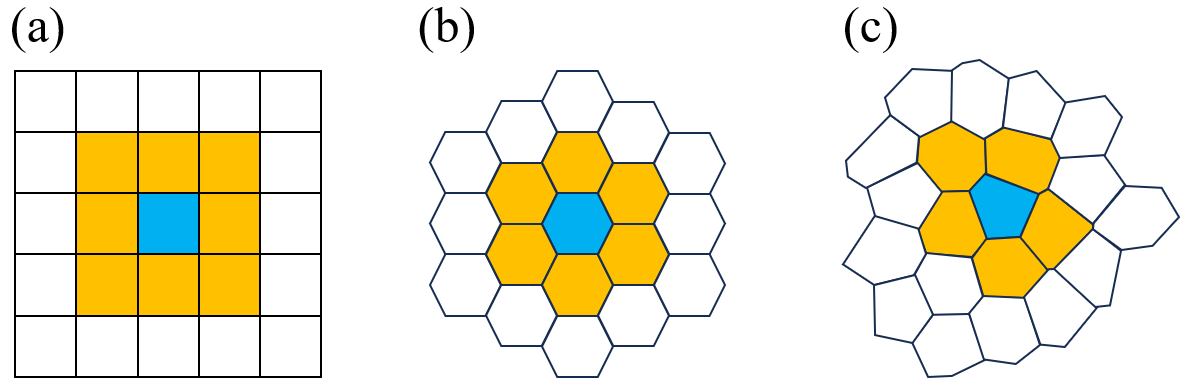}
\caption{\label{fgr:neighborhood}The definition of a neighborhood on different lattice types. (a) The Moore neighborhood on a square lattice, where the central site is blue and its 8 neighbors are indicated in yellow. (b) On the hexagonal lattice, there are 6 nearest neighbors to a central site. (c) On the irregular lattice, the sites having at least one shared vertex are neighbors.}
\end{figure}

When moving to irregular lattices, we will assume that for any lattice site $i$, we have $\mathcal{N}_{c}(i)=\mathcal{N}(i)$. In addition, we have assumed that the sites $i$ and $j$ are neighbors to each other, if and only if the polygons that contain these sites have at least one vertex in common. This simple rule on square lattice leads to Moore neighborhood which considers 8 neighbors for each site, see Fig.~\ref{fgr:neighborhood}a. On a hexagonal lattice, this leads to six neighbors, while the number of neighbors will vary per site on an irregular lattice, see Fig.~\ref{fgr:neighborhood}b,c, respectively. All of the neighborhoods considered thus far are in contact with the central cell. This is intuitive, as the interaction term considered in the CPM Hamiltonian describes surface tension. For discussion on other definitions of a neighborhood (for regular lattices) we refer to, for example, Refs.~\citenum{Durand2016, Magno2015}.

To work with an irregular lattice, the Hamiltonian of Eq.~\eqref{eq:hamiltonian_reg} needs to be modified. As the surface energy is proportional to the contact length of cells, we introduce a weight factor in the interaction term of the Hamiltonian. This weight ensures that the surface-energy penalty is proportional to the contact length of neighboring sites. The Hamiltonian then reads
\begin{align}
\label{eq:hamiltonian_vor} H &= \alpha \sum_{\langle i, j\rangle}
    w_{ij} \left( 1 - \delta_{\sigma(i),\sigma(j)} \right) + \lambda \sum_{\sigma=1}^{N_{c}} \left( a_{\sigma} - A_{0} \right)^{2} ,   
\end{align}  
in which the notation is mostly the same as in Eq.~\eqref{eq:hamiltonian_reg}. Here, the first summation on the right-hand side runs over all neighboring sites based on the aforementioned neighborhood, and $w_{ij}$ is the weight factor shared between sites $i$ and $j$. This weight is defined
\begin{align}
\label{eq:w_ij} w_{ij} &= l_{ij}/\Bar{l} ,
\end{align}
where $l_{ij}$ is the length of the contact edge between the sites $i$ and $j$, and $\Bar{l}$ is the average length of the edges taken over the entire lattice. Taking $w_{ij} = 1$, which is appropriate for regular lattices, the Hamiltonian of Eq.~\eqref{eq:hamiltonian_vor} reduces to that of Eq.~\eqref{eq:hamiltonian_reg}.

\begin{figure}
\centering
\includegraphics[width=\columnwidth,trim=8 25 8 9,clip]{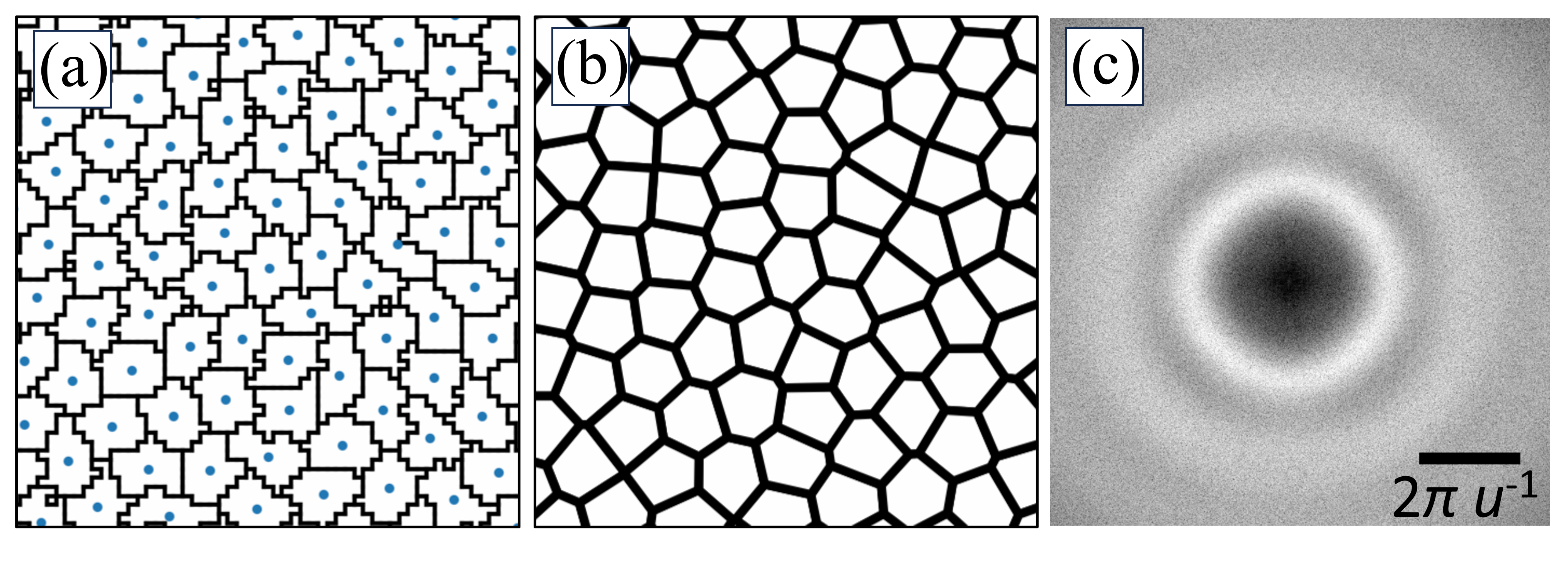}
\caption{\label{fgr:fluidity}Creation and characterization of the irregular lattice. (a) Part of a snapshot of the simulation, which was used to create a fluid state of cells. Nearly 0.1\% of all simulated cells are shown, outlined in black. The blue dots indicate the cell's centers of mass. (b) The Voronoi tessellation based on the snapshot in panel a. Each polygon will serve as a lattice site for our irregular lattice. (c) Fast Fourier transform of the picture of the center of mass of all cells comprising the fluid-like state --- a portion of which is shown in panel a. We express the inverse wavelength in terms of a length unit $u$, which represents the average cell spacing.}
\end{figure}

We generate our irregular lattices from a set of points using Voronoi tessellation. There are many ways in which to choose the generating points,~\textit{e.g.}, by choosing these randomly on the plane using a uniform distribution. However, this leads to the presence of many small Voronoi cells and several large ones\cite{Zhu2001, Kumar1993}. Here, we want our lattice sites to have roughly the same size and number of neighbors, whilst maintaining an isotropic character to the distribution of points. Therefore, we choose to base our lattice on the center of mass (CMS) of uniformly sized particles in a fluid phase. A regular CPM can be easily implemented and can serve to generate such a configuration for a suitably chosen value of $\alpha = 0.8$, which places the configuration in the fluid phase. We performed a large-scale simulation to obtain an equilibrated fluid, see Table~\ref{tbl:parameters} for our choices, by which we obtained approximately $200^{2}$ lattice centers. This is illustrated in Fig.~\ref{fgr:fluidity}a\footnote{For this study, we did not need to introduce any specific length unit. Lengths may be subsumed in the definitions of the prefactors. Only for the fast Fourier transforms, we introduce the length scale $u$, which makes the wave space vectors scale as $u^{-1}$.}.

Next, we applied Voronoi tessellation to these centers, imposing periodic boundary conditions, using the package Voro++~\cite{Rycroft2009}, see Fig.~\ref{fgr:fluidity}b. Finally, we verified that the newly formed irregular lattice does not have any long-range orientational structure. Using the image analysis software ImageJ~\cite{Schneider2012}, we performed a fast Fourier transform (FFT) on a snapshot of the center of mass of the lattice sites. The result is shown in Fig.~\ref{fgr:fluidity}c, which reveals the uniform rings, that are indicative of structural homogeneity. Further simulation details will be provided in Section~\ref{sub:parameters}. 

\subsection{\label{sub:parameters}Simulation Parameters}

The dynamics of the system is generated through MC attempts, as described in Section~\ref{sub:CPM}. We refer to $N_{s}$ MC attempts as a \textit{sweep} and we measure `time' in terms of MC sweeps (MCS). As the temperature here is merely a scaling of the cell membrane fluctuations, we set $k_{\mathrm{B}}T = 1$ throughout. The number of MCSs used depends strongly on the state point under consideration since the system features slow dynamics. Table~\ref{tbl:parameters} provides the relevant choices for both the equilibration (or waiting) time $t_{w}$ and the time over which we sampled $t_{s}$. We examined the evolution of total energy, the shape parameters, and the order parameter, both of which will be introduced shortly, to establish the appropriate value of $t_{w}$.

We wanted to evaluate the phase transition induced by the variation in the surface tension of the cells. Therefore, we assumed $\lambda$ to be constant and equal to $1.0$, while we changed $\alpha$ as the control parameter in the range of $\alpha \in [1.0,4.0]$ on all lattices to go from a disordered diffusive dynamics of the cells to an ordered arrested one. The value of $\alpha$ was changed in steps of $0.2$ and smaller steps of ($0.02 \sim 0.04$) near the transition point, as appropriate. In all cases, we used periodic boundary conditions and modeled nearly 1000 cells, see Table~\ref{tbl:parameters} for the details. For the irregular lattice, we used a simulation box of size $L_x = 200.11$ and $L_y = 197.08$, such that it is perfectly tileable by hexagons having a target area of $A_{0} = 40$. Every configuration studied was set up (and remained) confluent. We also performed several independent simulations to generate statistics. 

As the initial condition on square lattice, we considered a rectangular arrangement of cells, each of which has dimensions $5\times 8$. On the hexagonal and irregular lattices, we did the same for $\alpha\leq 2.2$, while taking an equilibrated snapshot (prepared at $\alpha=2.2$) for higher values of $\alpha$. The reason for this was the slow equilibration of the system in the high-$\alpha$ range. Again, we emphasize that we started sampling after the system was equilibrated and all the transient effects were decayed.

\begin{table*}[htb]
\footnotesize
  \caption{Parameter choices and details of the simulations carried out for this study. These include the size of the simulation boxes ($L_x \times L_y$), the number of lattice sites ($N_s$) and number of cells ($N_c$), the preferred cell area ($A_0$) and the values of surface tension ($\alpha$). Waiting and sampling times ($t_w$ and $t_s$), and the number of independent simulations for each value of $\alpha$ are presented in the final three rows, specified by $\alpha$. The right-most column presents the values for the square-lattice simulation, by which we generated the fluid state from which the irregular lattice was obtained.}
  \label{tbl:parameters}
  \begin{tabular*}{\textwidth}{@{\extracolsep{\fill}}ccccc}
    \hline
                         & square                   & hexagonal                  & irregular                  & irr. gen.                  \\
    \hline
    $L_x\times L_y$      & $200^2$      & $199.15\times 199.87$&$200.11\times 197.08$ & $1280^2$    \\
    $N_{s}$                & $4\times10^4$        & $39804$              & $40960$              & $1280^2$             \\
    $N_{c}$                & $1000$               & $1000$               & $986$                & $40960$              \\
    $A_0$                & $40$                 & $39.80$              & $40$                 & $40$                 \\
    $\alpha$             & $[1,4]$              & $[1,4]$              & $[1,4]$              & $0.8$                \\
    \hline
                         & $2\times10^5 , \alpha \in [1, 1.4]$ & $2\times10^5 , \alpha \in [1, 1.66]$ & $2\times10^5 \sim 7\times10^5 , \alpha \in [1,2.12]$ &                      \\
    $t_w$ (MCS)               & $2\times10^5  , \alpha \in [1.6,2.4]$ & $2\times10^5 \sim 1\times10^6, \alpha \in [1.68, 2.4]$ & $3\times10^5 \sim 1\times10^6 , \alpha \in [2.16,2.6]$ &    $2\times10^4$                  \\
                         & $3\times10^6 \sim 8\times10^6, \alpha \in [2.6,4]$ &  $1.2\times10^6, \alpha \in [2.6, 4]$ & $1\times10^6, \alpha \in [2.8, 4]$  &                      \\
    \hline
                         & $2\times10^5 \sim 1\times10^6, \alpha \in [1, 1.4]$ & $2\times10^5, \alpha \in [1, 1.66]$ & $2\times10^5 \sim 1\times10^6, \alpha \in [1,2.12]$ &                      \\
    $t_s$ (MCS)               & $2\times10^6 \sim 8\times10^6 , \alpha \in [1.6,2.4]$ & $2\times10^6 \sim 6\times10^6, \alpha \in [1.68, 2.4]$ & $1\times10^6 \sim 5\times10^6, \alpha \in [2.16,2.6]$ &   \_                   \\
                         & $2\times10^6, \alpha \in [2.6, 4]$ & $2\times10^6, \alpha \in [2.6, 4]$ & $1\times10^6, \alpha \in [2.8, 4]$ &                      \\
    \hline
          Number of simulations per each $\alpha$               & 50 & 20 & 20 &              1        \\
    \hline
  \end{tabular*}
\end{table*}

\subsection{\label{sub:charact}Characterization of the Results}

We characterize the outcomes of our simulations in several  ways. First, we compute the MSD, $\langle r^{2}(t) \rangle$, by extracting the CMS of each cell after equilibrating the system, based on equation \ref{eq:MSD}.
\begin{align}
\label{eq:MSD} \langle r^{2}(t) \rangle  = \langle (\mathbf{r}(t+t_w) - \mathbf{r}(t_w))^2 \rangle.
\end{align}
Here, $\mathbf{r}$ is the position vector for the center-of-mass of each cell and $t$ denotes `time'. Henceforth, we identify $t$ as the number of MCS\footnote{This definition is subject to the caveat that MCS can be understood to be proportional to the real time in a system, but they are not the actual time.}. The angle brackets indicate averaging over the cells and taking an ensemble average. We also fitted power-laws to the extracted MSDs to evaluate the diffusive behavior of the cells. Here, we focused on the long-time behavior only, which we found to follow
\begin{align}
\label{eq:beta} \langle r^{2}(t) \rangle &\propto t^{\beta} ,
\end{align}
where $\beta$ is the scaling coefficient. In practice, we determined $\beta$ from the slope of the MSD after taking the log of both the time and MSD. To calculate the effective diffusion coefficient $D_{\mathrm{eff}}$, we use
\begin{align}
\label{eq:D_eff}
\langle r^2(t)\rangle &= 4 D_{\mathrm{eff}} t ,
\end{align}
for the long-time behavior, whenever $\beta \gtrsim 0.95$. By using $D_\text{eff}$, we identified solid-like and fluid-like states of the tissue. We did this by fitting a polynomial function to the diffusion coefficient, and pinpointing where the curvature of the function is maximally negative. 
Diffusion coefficient has been used in similar studies for the identification of phase transition in tissues \cite{Bi2016}.

Second, we consider the local bond-order parameters to establish the degree of hexatic order. This is computed as follows for the cell indexed $k$
\begin{align}
\label{eq:psi} \psi_6(k) &= \frac{1}{N_{n}} \sum_{j \in \mathcal{N}(k)} e^{6 \iota \theta(j,k)} .
\end{align}
In this equation, $\mathcal{N}(k)$ is the set of nearest neighbors to the $k$-th cell and $N_{n}$ is the number of neighbors. Here, $\mathcal{N}(k)$ is considered simply six nearest neighbors\footnote{Other definitions of the local hexatic bond order are possible~\cite{Digregorio2018, borba2013}, but we considered the definition provided in Eq.~\eqref{eq:psi} the most appropriate for our purposes.} of the cell $k$. This means that $N_{n}=6$ for all the cells. The angles $\theta(j,k)$ represent the counterclockwise angle between the $x$-axis and the vector connecting the center-of-mass of the cell $k$ to that of $j$. The $\iota$ represents the complex identity, $\iota^{2} = -1$. For a perfect hexagonal arrangement, the absolute $\vert \psi_{6}(k) \vert = 1$, and the value of the hexatic order decreases with disorder. Typically, we average $\vert \psi_{6} \vert$ over all cells and the production time.

\begin{figure}[!htb]
\centering
\includegraphics[width=\columnwidth,trim=20 1 1 7,clip]{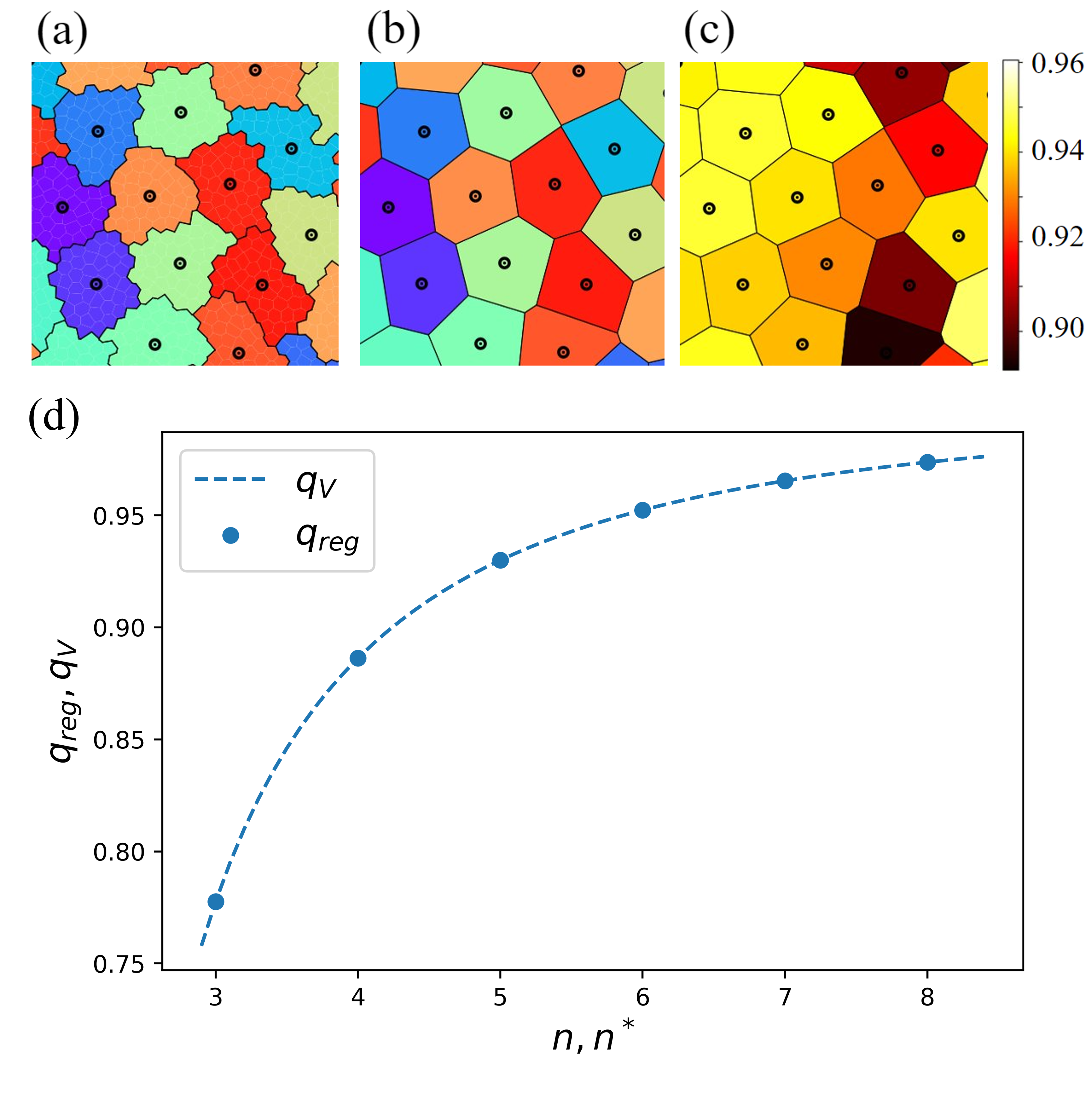}
\caption{\label{fgr:Vor_tes_N_star}Using Voronoi tessellation to determine the shape index of lattice-based cells. (a) A snapshot of cells on the disordered lattice. The colors indicate individual cells and the black circles their centers of mass. (b) Voronoi tessellated version of the cells based on the position of the centers of mass. (c) Voronoi tessellated cells are colored by their isoperimetric quotient $q_{\mathrm{V}}$, as indicated by the color bar shown on the right-hand side. The data derives from a part of a simulation with $\alpha=1.8$. (d) Isoperimetric quotient of regular polygons, $q_{\mathrm{reg}}(n)$,  as well as its generalization $q_{\mathrm{V}}(n^*)$.}
\end{figure}

Third, we complement the hexatic-order analysis by considering the dimensionless quantity called the \textit{shape index}. For any 2D shape, this quantity is defined as $P/\sqrt{A}$, where $P$ and $A$ are the perimeter and the area, respectively. This parameter has been studied extensively in tissue mechanics,~\textit{e.g.}, see Refs.~\citenum{Bi2016, Park2015, Bi2015, Hertaeg2024, Krajnc2020, Chiang2016, Thomas2023, Li2023}. A slightly modified version of this quantity is the \textit{isoperimetric quotient}, which is defined as $q = \sqrt{4 \pi A}/ P$. It is dimensionless as well, and equal to 1.0 for the circle, while being close to 0 for highly elongated shapes.

Here, we should note that the cell perimeter derived from lattice-based models is not readily comparable to that in off-lattice models. 
This is because the borders of the cells are defined by the edges of the lattice sites, which enforces excess jaggedness to the cells --- in mathematical language, a natural metric to distance calculations. To overcome this issue, we applied Voronoi tessellation to the center-of-mass (CMS) positions of the cells. This enabled us to study the effective areas and perimeters of the generated polygons and compute the associated isoperimetric quotient. 
Note that the effective cell shape obtained by applying Voronoi tessellation can be different from the original. There are alternative ways to smoothen the jagged borders of cells on a lattice,~\textit{e.g.}, using elliptic Fourier analysis~\cite{kuhl1982, tweedy2013, sanchez2018}. However, since one of the goals of this study is to compare the shape characteristics in CPM with those in vertex and Voronoi models, we decided to construct our effective polygons in the same manner.
We use the subscript `V', for example, $q_{\mathrm{V}}$, to indicate the use of our Voronoi procedure. 
Figure~\ref{fgr:Vor_tes_N_star}(a-c) shows a zoomed-in view of a snapshot that illustrates the process, as well as the obtained value of $q_{\mathrm{V}}$. We should note that a correction was also introduced to evaluate the perimeter of the cells on the CPM~\cite{Magno2015, rens2019}. However, we chose to use Voronoi tessellation to make a clear comparison between this study and the vertex and Voronoi models that are being used in this context of epithelia.

The isoperimetric quotient of regular polygons with $n$ edges, $q_{\mathrm{reg}}(n)$, can be readily computed and reads
\begin{align}
\label{eq:nast} q_{\mathrm{reg}}(n) &= \frac{ \sqrt{4\pi A_{\mathrm{reg}}(1,n)} }{ P_{\mathrm{reg}}(1,n) } = \frac{ \sqrt{2n\pi\sin{(2\pi/n)} } }{ 2n\sin{(\pi/n)} } ,
\end{align}
where $A_{\mathrm{reg}}(r,n)$ and $P_{\mathrm{reg}}(r,n)$ are the area and the perimeter of a regular polygon having $n$ edges and a circumscribing circle with radius $r$. This function maps the number of the edges of regular polygons to their isoperimetric quotient, and \textit{vice versa}. We can now straightforwardly apply the right-hand side of Eq.~\eqref{eq:nast} for non-integer values of $n$, which we call the \textit{generalized edge number} and which we will identify using $n^{\ast}$. Numerically inverting this function, we uniquely obtain $n^{\ast}$ for any given value of $q_{\mathrm{V}}$. By doing so, we can assess the effective number of edges for a given (convex) cell shape. The dependence is shown in Fig.~\ref{fgr:Vor_tes_N_star}d.

Lastly, we also considered the \textit{circularity} $C$ of our model cells. This parameter was introduced by Zunic and Hirota~\cite{uni2008, uni2010}, and is calculated as follows. Given the area $A$ of a 2D connected object, and the moment of inertia tensor $\bar{I}$, we write
\begin{align}
\label{eq:circ} C &= \frac{1}{2\pi} \frac{ A^{2} } { \bar{I}_{xx} + \bar{I}_{yy} } ,
\end{align}
where in the denominator, the trace of the tensor is taken. Since the trace is invariant under translation and rotation, $C$ is independent of the coordinate system. 
One should be careful in connection to Refs.~\citenum{uni2008} and~\citenum{uni2010},  to note that people also use the term `circularity' to refer to the quantity $4\pi A^2/P$, which is $q^2$ in our characterization. However, throughout this study, by circularity, we mean the quantity calculated using equation \ref{eq:circ}. 
The reason we study this parameter is that, unlike the isoperimetric quotient, it is not directly dependent on the perimeter. As we have indicated above, there are issues in defining a perimeter length in lattice-based models. Circularity bypasses this issue and thus helps us learn how different ways of measuring cell roundness compare to each other.
Circularity is in fact, one of the Hu moment invariants~\cite{MingKueiHu1962} that are widely used in image processing and pattern recognition~\cite{Nasrudin2021}.
This quantity has also been used in studying cell morphology~\cite{Kopanja2018}, branching patterns in organogenesis simulations~\cite{Poorkhanalikoudehi2021}, and pathological cell nucleus shape analysis~\cite{Ammar2017}. 
However, to the best of our knowledge, thus far it has not been studied in the context of confluent tissue transitions. Similar to the isoperimetric quotient, we rely on a Voronoi tessellation to determine $C_{\mathrm{V}}$. The circularity is much more strongly nonlinear than $q$ in the number of edges of a regular polygon. Therefore, we do \textit{not} invert it to establish an equivalent circularity-derived generalized edge number $n^{\circ}$.

\section{\label{sec:results}Results}

In this section, we introduce the main results of our CPM simulations. As explained in Section~\ref{sub:CPM}, the control parameter in our simulations is $\alpha$, which models the surface tension between cells. We start by showing that increasing $\alpha$ leads to a first-order phase transition from a disordered fluid to an ordered hexagonal phase on all lattices.

\subsection{\label{sub:hex}Hexatic Order, Artifacts, and Diffusion}

\begin{figure*}[!htb]
\centering
\includegraphics[width=\textwidth,trim=5 5 5 5,clip]{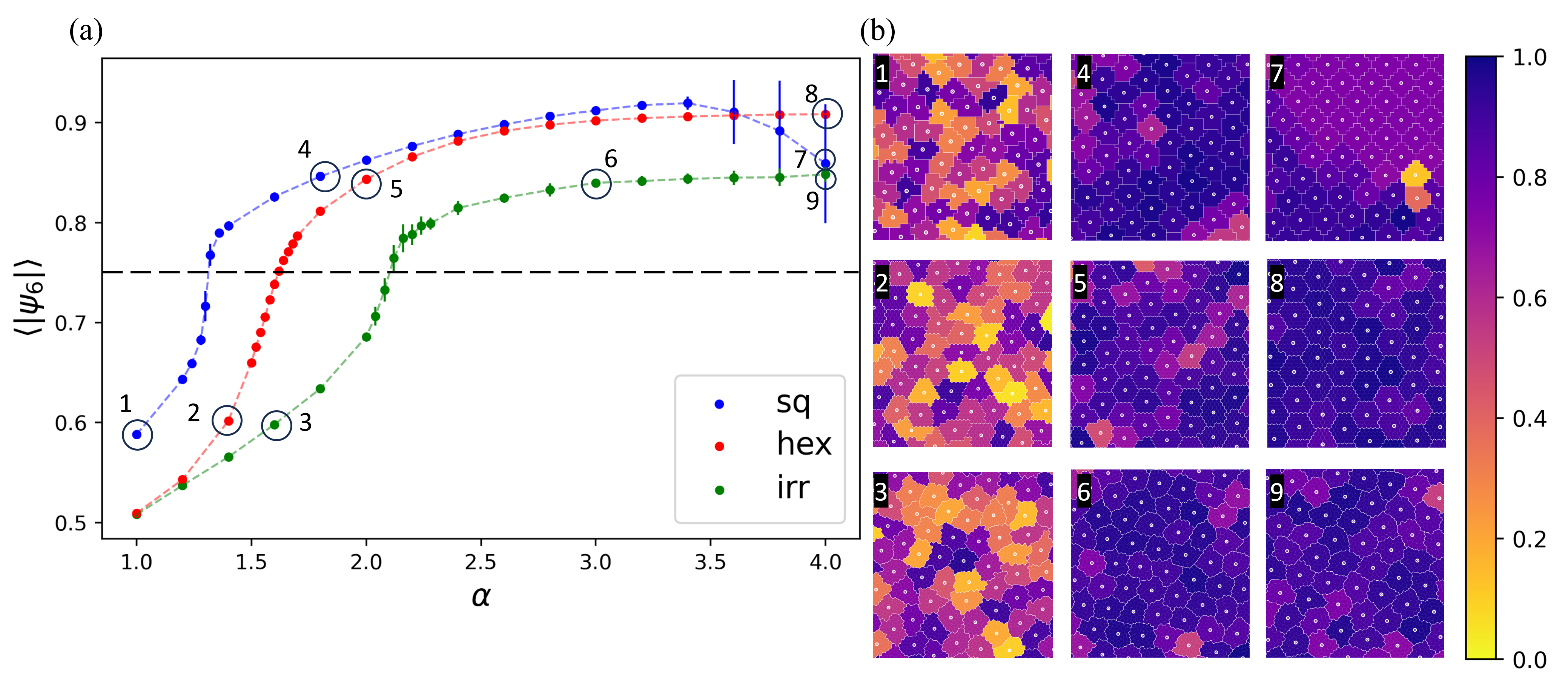}
\caption{\label{fgr:Phase_transition_hex}The formation of hexatic order on the three lattice types considered by increasing the surface tension. (a)~The system-averaged hexatic order $\langle \vert \psi_{6} \vert \rangle$ as a function of $\alpha$ for the CPM on a square (blue), irregular (green), and hexagonal (red) lattice. The error bars show the standard error of the mean. The black dashed line indicates the transition value reported by Ref.~\citenum{Durand2019}. (b) Snapshots of the simulations specified by their labels in panel (a); only a small part of the simulation box is shown ($\sim 6 \%$). The color of the cells indicates their local hexatic order.}
\end{figure*}

Figure~\ref{fgr:Phase_transition_hex} shows the (average) hexatic orientational order parameter $\langle \vert \psi_{6} \vert \rangle$ as a function of $\alpha$ and several representative snapshots of a part of the simulation area. The average hexatic order is generally an increasing function of $\alpha$, except for high values of $\alpha$ on the square lattice, which we will return to shortly. For all our lattices there is a substantial increase in $\langle |\psi_6| \rangle$ in the range 0.7 to 0.8, which appears to connect a disordered and an ordered `branch' in the state space. This is indicative of the presence of a first-order phase transition from a disordered fluid to a hexagonal solid. The nature of the structural change is further confirmed by contrasting the first two columns of snapshots in Fig.~\ref{fgr:Phase_transition_hex}b with each other. The centers of mass of the cells assume a hexagonal arrangement in the three subpanels.

In the case of a square lattice, the transition appears sharp. For the other two lattices, there is a smoother transition, which can be attributed to the fact that we work in the $NVT$ ensemble. That is, when we average over several realizations of the system, these likely contain both fluid and hexagonal configurations, beyond the transition $\alpha$. 
In support of this, we found patterns of coexistence between the disordered and ordered phases in some of the simulations, see Fig.~S1 for an example snapshot. This provides further (tentative) evidence for the presence of a first-order transition. Note that unlike in fluid-solid phase coexistence for particle-based systems~\cite{kirchner2024, akiyama2017, camley2016, basan2013, campo2019}, we have an area-constraint term in our Hamiltonians~\eqref{eq:hamiltonian_reg} and~\eqref{eq:hamiltonian_vor}. This presumably narrows any density gap that could be present between the disordered and ordered phases, and makes it difficult to find strong evidence of coexistence.
We will also not concern ourselves with the existence of a potential intermediate hexatic phase here, but referencing the literature~\cite{Durand2019, Pasupalak2020, Li2021}, it should be present.

Increasing $\alpha$ beyond the transition value, $\langle \vert \psi_{6} \vert \rangle \approx 0.76$, we find that the hexatic order increases further.  Note that our transition value is  comparable to that reported in Ref.~\citenum{Durand2019}, providing additional confidence in our results. On a square lattice, $\langle \vert \psi_{6} \vert \rangle$ eventually assumes a maximum. We can appreciate the underlying cause of the high-$\alpha$ reduction by examining snapshot 7 in Fig.~\ref{fgr:Phase_transition_hex}b. This reveals that for the square lattice, the cells become distorted into a configuration with staircase-like borders. 
That is, the borders of cells in large parts of the simulation locally follow the lattice (are at 90$^{\circ}$ angles). This leads to the cell edges overall being directed at angles $\pm 45^{\circ}$ with respect to the horizontal (or vertical) and the cells assuming a rhombus-like shape. This, in turn, gives rise to the strong, unphysical peak in the distribution of hexatic order, as can be seen in Fig.~S3.
For the disordered lattice, there is no such a peak, and the distribution of hexatic order is smooth within the error, as can be appreciated from Fig.~S3b. Together these results underpin that on the disordered lattice, the artificial (rhomboid) cell shapes are not present.

The orientation of the cell borders on the hexagonal lattice is also a lattice artifact. Figure~\ref{fgr:Phase_transition_hex}b, snapshots~5 and~8, show that the borders of neighbor cells at intermediate-to-high surface tension, follow specific directions. These preferential orientations are dictated by the hexagonal symmetry of the underlying lattice. The effect appears less `severe' than for the cells on the square lattice, where artifacts lead to significant cell-shape distortions. However, we caution against drawing this conclusion, as the natural high-surface-tension shape of the cells is hexagonal, which masks the extent to which cells are constrained by the underlying lattice.

\begin{figure}[!htb]
\centering
\includegraphics[width=\columnwidth,trim=1 1 1 1,clip]{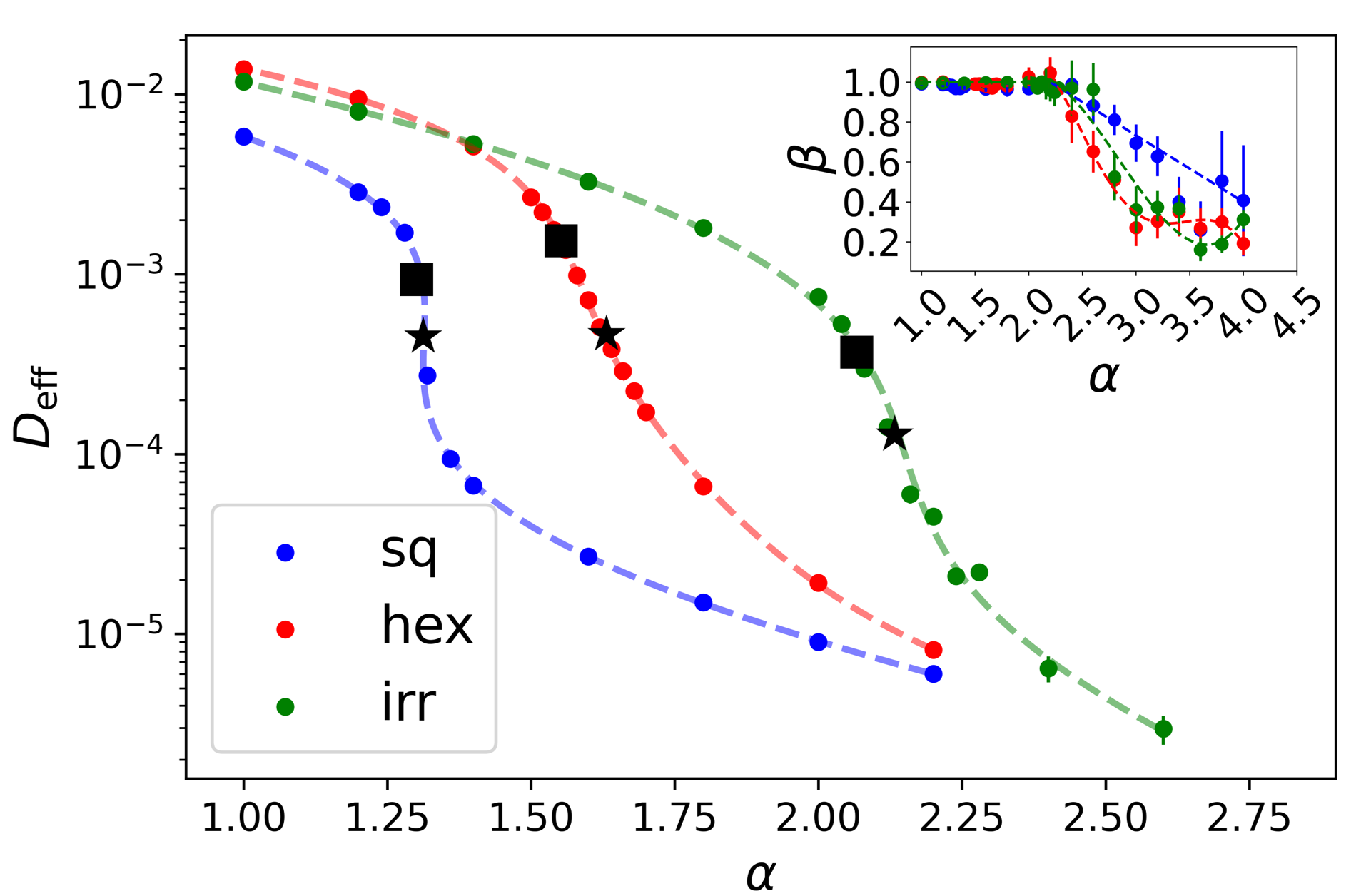}
\caption{\label{fgr:beta_diff_a_b}
Particle diffusion indicates where the system transitions from fluid-like to solid-like behavior. The effective diffusion coefficient $D_{\mathrm{eff}}$ of cells as obtained from their long-time diffusive dynamics as a function of the interfacial energy $\alpha$ for the three different lattices as labeled. We only show data for which the scaling exponent of the MSD, $\beta \approx 1.0$, and the (long-time) dynamics is indeed diffusive. The inset shows $\beta$ as a function of $\alpha$. The dashed lines in the main panel and the inset serve as guides to the eye. 
The star symbols ($\star$) localize  the inflection points to fitted data, and the square symbols ($\blacksquare$) indicate 
the transition points which were determined from the second derivative of the fitted data.
See the main text for the procedure.
}     
\end{figure}

We further characterized the behavior of our systems by examining the MSD scaling exponent $\beta$, and the diffusion coefficient, $D_{\rm{eff}}$, as defined in Eq.~\eqref{eq:beta} and~\eqref{eq:D_eff}, respectively. 
A plot of MSD on the disordered lattice is shown in Fig.~S2. 
First, we evaluated $\beta$ and $D_{\rm{eff}}$ by fitting lines to log-log plots of the late-time MSD. Figure~\ref{fgr:beta_diff_a_b} shows both quantities for different lattices as a function of $\alpha$ ($\beta$ is shown in the inset). 
For low values of $\alpha$, we readily obtain diffusive scaling $\beta = 1$ within the error, and the system is in a fluid-like state. 
Since the definition~\eqref{eq:D_eff} holds for $\beta = 1$, we only show $D_{\rm{eff}}$ for the cases where $\beta$ departs from $1$ by less than one standard error of the mean. Note that the decrease in $D_{\mathrm{eff}}$ can be in orders of magnitude over the entire range of considered $\alpha$, though we do not consider this indicative of glassy behavior, as reported in other sources~\cite{Chiang2016, Bi2016, Sadhukhan2021}; we will return to this point in Section~\ref{sec:discuss}. For high values of $\alpha$ we observed sub-diffusive behavior. It is likely that for some of these values, the dynamics eventually becomes diffusive when they are evaluated for a sufficiently long time. However, we did not test this, as these values of $\alpha$ are far enough away from the transition that they do not affect our analysis and conclusions.

Figure~\ref{fgr:beta_diff_a_b} reveals that $D_{\rm{eff}}$ generally decreases with $\alpha$. It is difficult to identify where the phase transition happens without performing an exhaustive analysis. Here, we therefore computed two properties. First, the decrease has an inflection point at a given value of $\alpha$ that is lattice-dependent. This inflection point is posited to be indicative of coexistence,~\textit{i.e.}, it is caused by blending fluid-like and solid-like behavior in equal parts. Thus, we expect the inflection point to overestimate the value of $\alpha$ for which there is a transition. To locate the inflection $\alpha$, we fitted polynomial curves to the semi-log plot of $D_{\rm{eff}}$. 
The obtained points are indicated using stars ($\star$) in Fig.~\ref{fgr:beta_diff_a_b}. We will refer to the associated values using $\alpha^{\star}$ and $D^{\star}_{\rm{eff}}$, respectively. Second, we identify the point where $D_{\rm{eff}}$ starts to drop rapidly by examining the second derivative of the fitted $D_{\rm{eff}}(\alpha)$. 
To do so, we considered the point at which, the curvature of the fitted curve is maximally negative. 
This point is posited to be closer to the transition,~\textit{i.e.}, at the end of the purely disordered branch. We denote this point using $\alpha^{\blacksquare}$ and $D^{\blacksquare}_{\rm{eff}}$ which are showed by square symbol if Fig.~\ref{fgr:beta_diff_a_b}.

\begin{figure}[!htb]
\centering
\includegraphics[width=\columnwidth,trim=2 2 2 8,clip]{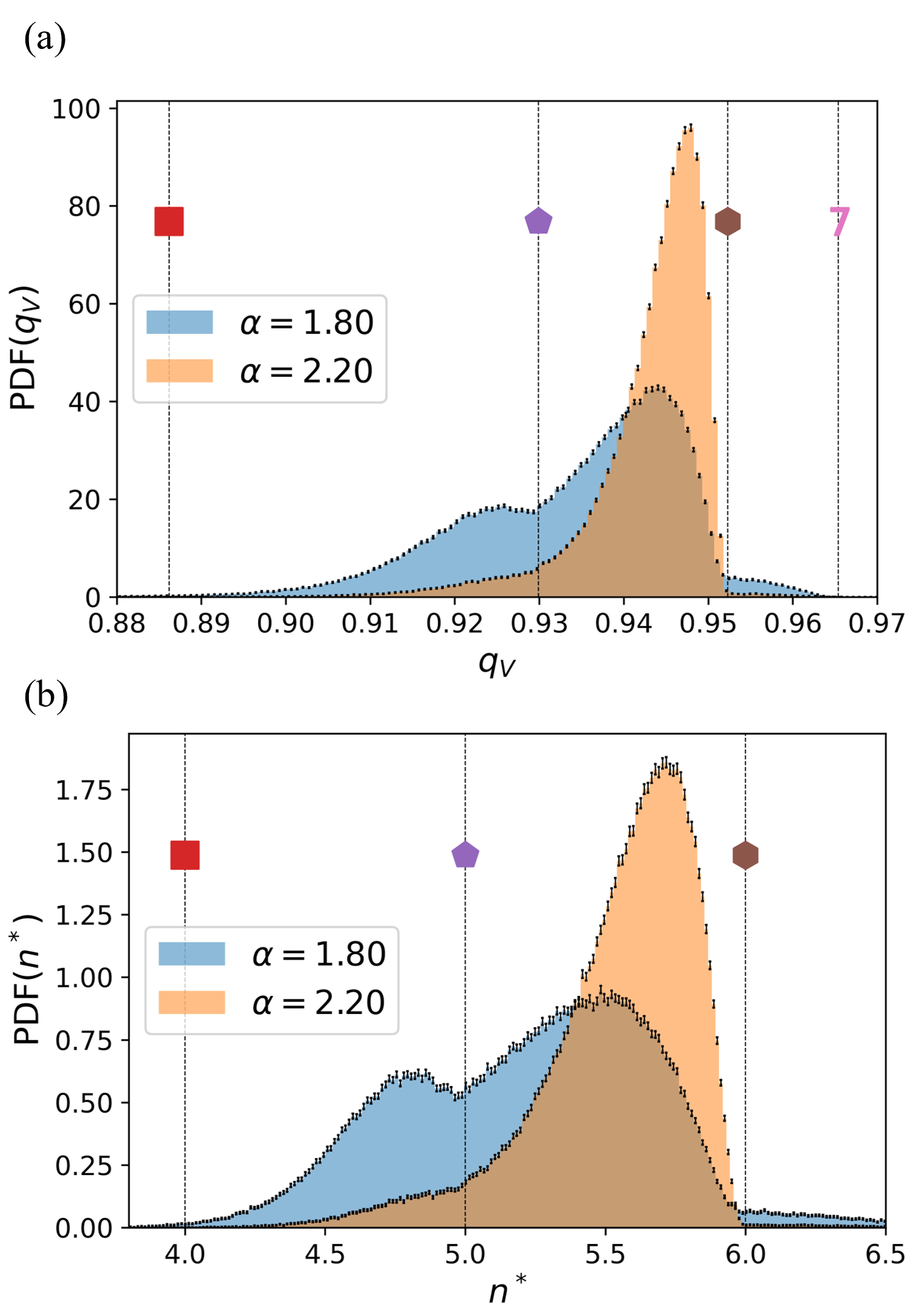}
\caption{\label{fgr:q_V_N_star_dist_ex}The distribution of Voronoi-based isoperimetric quotient $q_{\mathrm{V}}$ for a liquid- and solid-like state. (a) For a disordered lattice, we computed the probability density function (PDF) of $q_{\mathrm{V}}$ for $\alpha=1.8$ (blue) and $\alpha = 2.2$ (orange), which are in the fluid and crystal states, respectively. (b) The same data, but now converted to an effective number of edges $n^{\ast}$. In both graphs, the error bars indicate the standard error of the mean, and values of $q_{\mathrm{V}}$ and $n^{\ast}$ that correspond to regular polygons are indicated using dotted vertical lines and the use of symbols/numbers.}
\end{figure}

\subsection{\label{sub:iso}Isoperimetric Quotient and Circularity}

Figure~\ref{fgr:q_V_N_star_dist_ex}a shows the equilibrium distribution of the Voronoi-based isoperimetric quotient, $q_{\mathrm{V}}$, for two different values of $\alpha$, obtained by using an irregular lattice. Because our model tissue is confluent, the $q_{\mathrm{V}}$ for the majority of the cells assumes values in the square-to-hexagon range. Note that for the fluid-like state, there is a dip in the value $q_{\mathrm{V}}$ around that of a regular pentagon. This is a consequence of the definition of $q_{\mathrm{V}}$, rather than a profound result. Voronoi cells with $5$ vertices always have values of $q_{\mathrm{V}} \leq q_{\mathrm{reg}}(5)=0.930$, and cells with $6$ vertices always have $q_{\mathrm{V}} \leq q_{\mathrm{reg}}(6)=0.952$.
Whether the role of an arbitrary polygon in tiling the plain is similar to that of a pentagon or a hexagon, depends mostly on whether its $q_{\mathrm{V}}$ is closer to $q_{\mathrm{reg}}(5)$ or $q_{\mathrm{reg}}(6)$, rather than the number of its edges. 
It is clear that upon undergoing a phase transition, the distribution shifts directly from being double-peaked to being strongly peaked close to a hexagonal value --- another indicator of a first-order phase transition. The fact that the model tissue has defects in combination with the properties of $q_{\mathrm{V}}$, makes it that the peak is not exactly centered about $q_{\mathrm{V}} \approx 0.952$. The defects can be seen in Fig.\ref{fgr:Phase_transition_hex}b, snapshots~6 and~9.

The distribution of $q_{\mathrm{V}}$ is insightful, but to help understand the properties of the underlying system, it is beneficial to convert it to an effective number of edges $n^{\ast}$, see Fig.~\ref{fgr:q_V_N_star_dist_ex}b and the definition in Section~\ref{sub:charact}. The advantage of working directly with $n^{\ast}$, rather than $q_{\mathrm{V}}$, is that it allows us to examine the partition of the distribution. We identify the nearest integer number to a given $n^{\ast}$ by $\lfloor n^* \rceil$, which factors our range into distinct subsets of triangular, square, pentagonal, hexagonal, heptagonal,~\textit{etc.} neighborhoods. For example, all the cells having $\lfloor n^{\ast} \rceil = 5$ (\textit{i.e.}, $4.5 < n^{\ast} \leq 5.5$) can be referred to as \textit{pseudo-pentagons}. Integrating the PDF belonging to the pseudo-pentagons provides insight into the fraction of pentagonal cells in the system. This includes cells with 5 or more edges, but for which some of the edges are very small.

\begin{figure}[!htb]
\centering
\includegraphics[width=\columnwidth,trim=4 2 4 10,clip]{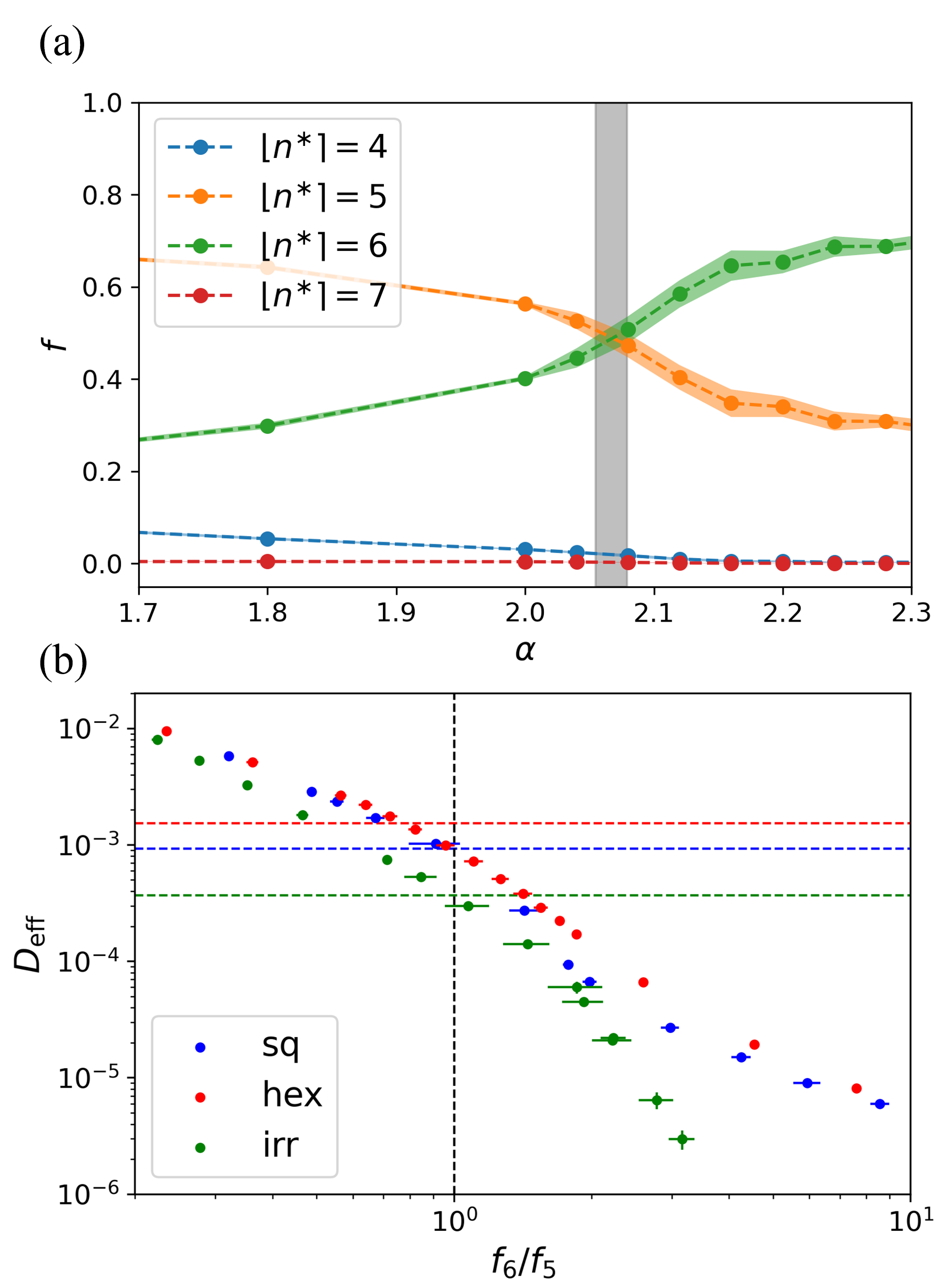}
\caption{\label{fgr:N_star_fractions_a_b}The trend in the fractions of pseudo-polygons can be used to determine the transition point. (a)~The fractions of pseudo-polygons $f_{n^{\ast}}$ as a function of the surface tension $\alpha$ obtained for a disordered lattice. The number of edges for the pseudo-polygons is indicated in the legend and the colored area around the respective curves indicates the standard error of the mean. The vertical gray bar shows the range for which the system transitions from a fluid to a hexagonal solid; \textit{i.e.}, where the diffusion coefficient drops steeply. (b)~The diffusion coefficient $D_{\rm{eff}}$ as a function of the ratio $f_{6}/f_{5}$. The horizontal dashed lines show $D^{\blacksquare}_{\rm{eff}}$ for different lattices. The vertical dashed line highlights the crossover at $f_6/f_5=1$.}
\end{figure}

Figure~\ref{fgr:N_star_fractions_a_b}a shows fractions of pseudo-polygons $f_{n^{\ast}}$ present in the system as a function of $\alpha$. Interestingly, the curves for $f_{5}$ and $f_{6}$ show a crossover that matches well where we locate the phase transition based on our analysis of the diffusion coefficient drop. The physical intuition in this representation is clear: to crystallize, the system must have a majority of pseudo-hexagons. Figure~\ref{fgr:N_star_fractions_a_b}b shows the relation between the crossover and the transition on different lattices. Here, we have plotted $D_{\rm{eff}}$ as a function of $f_{6}/f_{5}$.  
The transition value of diffusion coefficient, i.e., $D^{\blacksquare}_{\rm{eff}}$, is indicated by the horizontal dashed lines.
In Fig.~\ref{fgr:N_star_fractions_a_b}b, it can be seen that when the ratio exceeds 1, the diffusion coefficient drops sharply, which is a precursor to full crystallization, on both the square and irregular lattices. 
However, on the hexagonal lattice, there is a slight mismatch between the crossover and the transition. The supplemental information goes into a longer discussion of this crossover for regular lattices and, also considers the mean and median of the distribution. The latter quantity also appears to be a good quantifier of the transition for these systems, see Figs.~S4 and~S5.

\begin{figure}[!htb]
\centering
\includegraphics[width=\columnwidth,trim=1 1 1 1,clip]{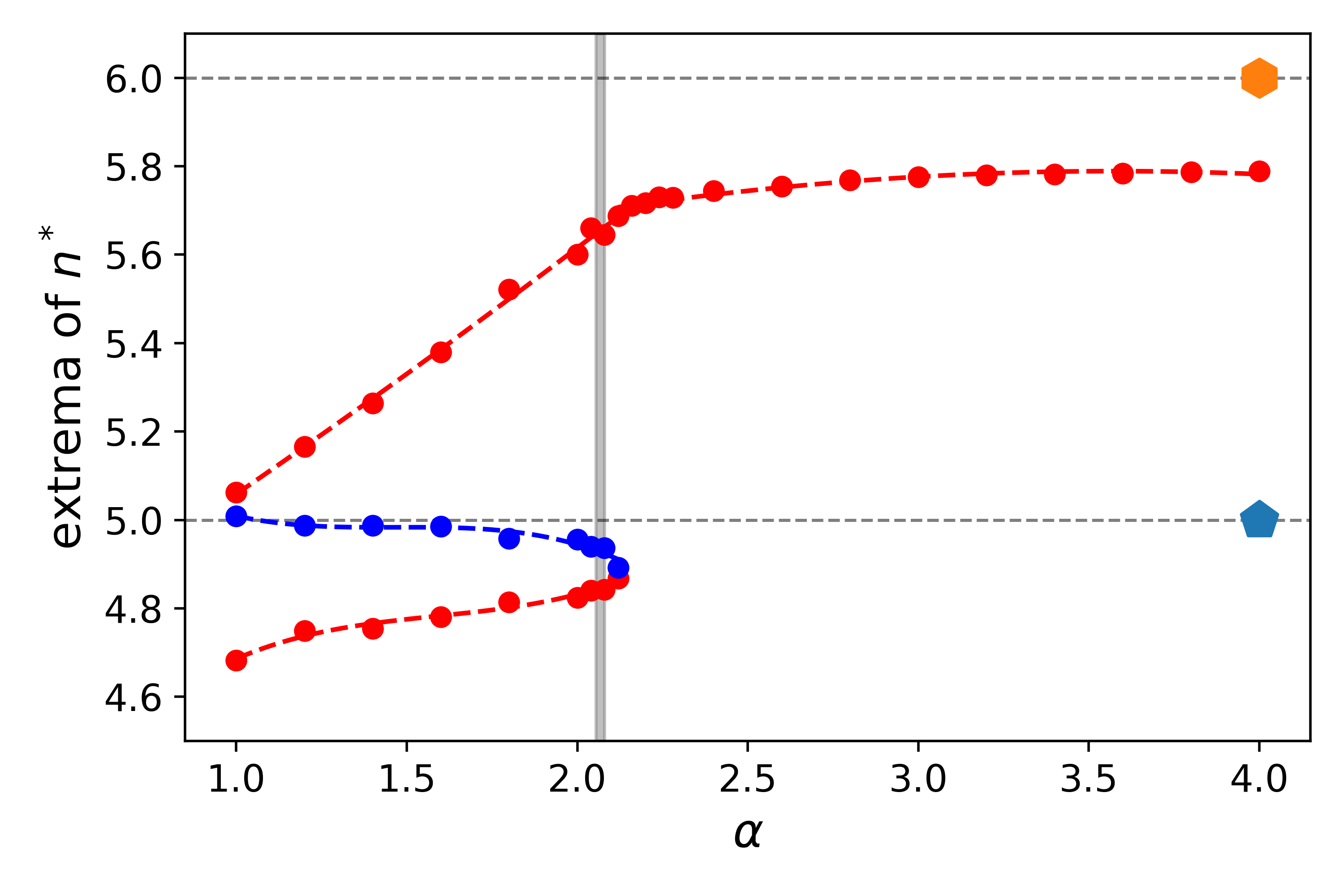}
\caption{\label{fgr:N_star_extrema}The fluid-to-solid transition in the model cell system as characterized using the extrema in the distribution of the Voronoi-based isoperimetric quotients. The $n^{\ast}$ values for the local maxima are indicated in red and the local minimum in blue when it is present. The dashed lines are guides to the eye and the grey vertical bar indicates $\alpha$ interval in which we locate the transition, similar to Fig.~\ref{fgr:N_star_fractions_a_b}a. The dotted horizontal lines indicate the values for a regular pentagon and hexagon, respectively, as labeled. The error bars are smaller than the point size. All data was obtained for a disordered lattice underlying the CPM dynamics.}
\end{figure}

Returning to the data presented in Fig.~\ref{fgr:q_V_N_star_dist_ex}, we see that the distributions for the fluid-like state (\textit{e.g.}, $\alpha=1.8$) have two peaks, while the solid-like state (\textit{e.g.}, $\alpha=2.2$) is characterized by a single peak. This statement holds in general and we can extract the positions of $n^{\ast}$ for the maximum; or the two maxima and the local minimum, see Fig.~\ref{fgr:N_star_extrema}. Similar data for the square and hexagonal lattices are provided in Fig.~S6. We note that the behavior of the peak bears the hallmarks of a first-order phase transition, with a jump in the value of the `effective order parameter' at $\alpha \approx 2.1$, as we alluded to before. The minimum is effectively at $n^{\ast} = 5$, as is a feature of the definition of $q_{\mathrm{V}}$. When the system transitions into the solid state, the peak reaches a nearly constant value of $n^{\ast} \approx 5.7$, which is shared among the studied underlying lattices. This is commensurate with the model cells transitioning to a state that is geometrically similar between the square, irregular, and hexagonal lattices. However, we emphasize that this does not imply that the transition is unaffected by the properties of the underlying lattice, as we have provided evidence for above.

\begin{figure}[!htb]
\centering
\includegraphics[width=\columnwidth,trim=10 1 1 10,clip]{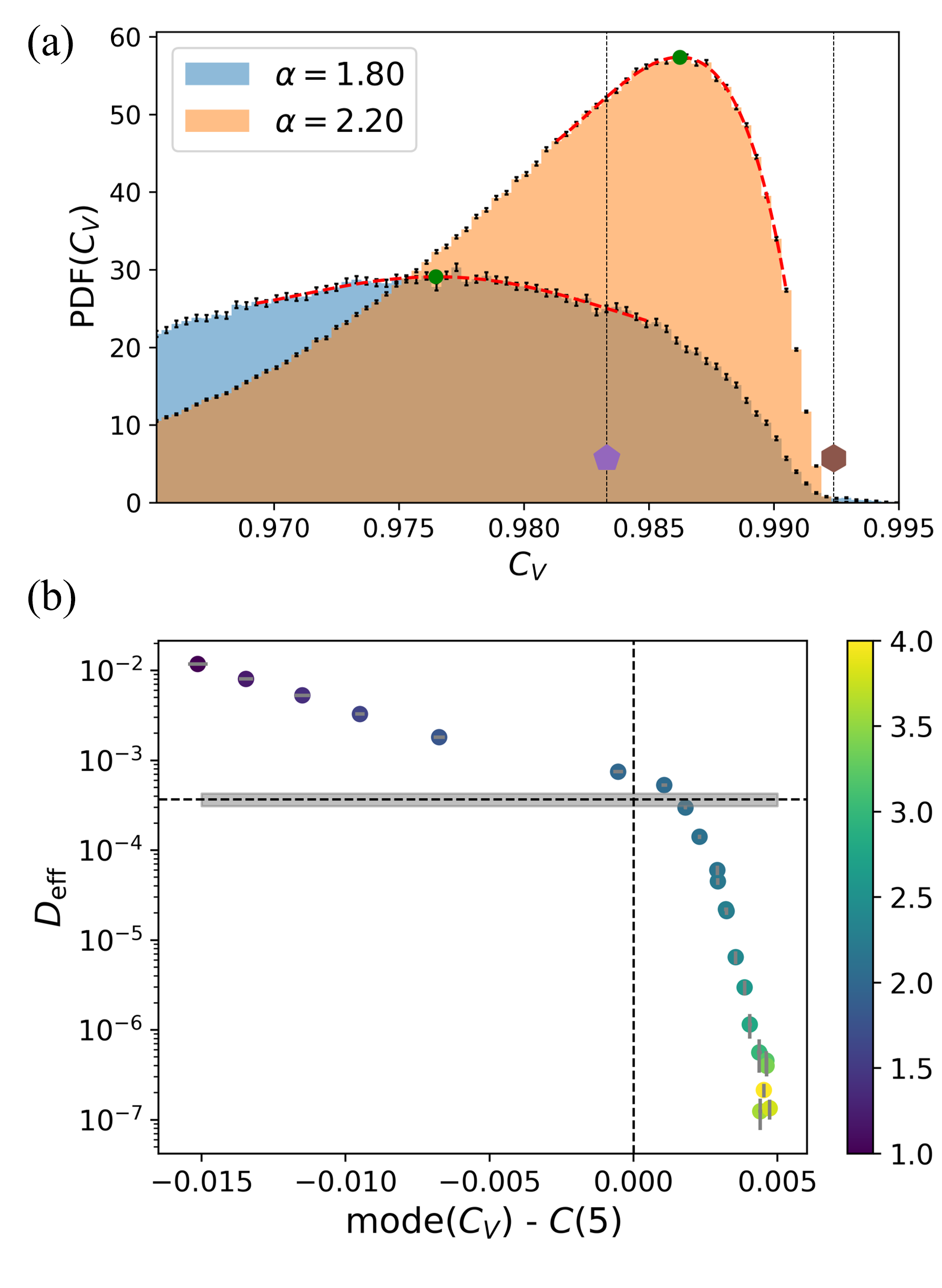}
\caption{\label{fgr:cV_mode_irr}The circularity $C_{\mathrm{V}}$ of the Voronoi tessellation as an indicator of the phase transition. (a) The PDF of $C_{\mathrm{V}}$ for two different values of $\alpha$ on the disordered lattice. One is representative of the fluid-like regime ($\alpha=1.8$, blue), and the other, of the solid-like regime ($\alpha=2.2$, orange). The position of the peak of the distribution (green circles) is extracted from the red (polynomial) fits. The two dotted vertical lines and symbols indicate the value of circularity for a regular pentagon and hexagon, respectively. (b) The diffusion coefficient, $D_{\rm{eff}}$  as a function of the departure between the mode of the distribution, $\mathrm{mode}(C_{\mathrm{V}})$, and the value of the circularity for a regular pentagon $C(5)$. 
The horizontal dashed line, and the gray box, show the average and standard error of $D^{\blacksquare}_{\rm{eff}}$.
The data points are colored by the value of $\alpha$, as indicated in the color bar to the right. In both panels, the error bars show the standard error of the mean.}
\end{figure}

Lastly, we calculated the circularity of the Voronoi-generated polygons for our model tissues. Figure~\ref{fgr:cV_mode_irr}a shows the distribution of $C_{\mathrm{V}}$ of the cells on the disordered lattice, for two values of $\alpha$ already considered in Fig.~\ref{fgr:q_V_N_star_dist_ex}. It turns out that the distributions of $C_{\mathrm{V}}$ for all the values of $\alpha$ studied here are unimodal, unlike those of $q_{\mathrm{V}}$,~\textit{e.g.}, see Figs.~\ref{fgr:cV_mode_irr} and~S7. We also observe from our data that at the transition point, the mode of the distribution of $C_{\mathrm{V}}$ almost matches with the circularity of a regular pentagon, which we identify with $C(5)$. If we apply the same strategy as we did to calculate $n^\ast$ from $q_{\mathrm{V}}$, to calculate $n^\circ$ from $C_{\rm{V}}$, the mode of the distribution lies around $n^\circ \approx 5.1$.
This relation is shown in Fig.~\ref{fgr:cV_mode_irr}b for the disordered lattice, where we calculated $D_{\rm{eff}}$ for the long-time diffusive regime, as a function of the departure of the mode from the pentagonal value,~\textit{i.e.}, $\mathrm{mode}(C_{\mathrm{V}}) - C(5)$. Clearly, the value $C(5)$ is an excellent estimator for the phase transition. Lastly, it should be reemphasized that circularity compares the area with moment of inertia. This gives it an advantage over the isoperimetric quotient, which depends on perimeter length.

\section{\label{sec:discuss}Discussion}

In this section, we provide additional context to our research. We have split this into five parts for convenience. The first concerns itself with the quality of our phase diagram and the lack of a glassy phase. The second makes the connection to experimental observations to further justify the use of the CPM. The third discusses resolving the lattice artifacts present on regular lattices. The fourth provides insight into the efficiency of our algorithm and the challenges in working with an irregular-lattice CPM. We close with a larger discussion on the use of shape parameters.

\subsection{\label{sub:phase_glass}The Phase Transition}

On three different lattices studied here, we observed an order-disorder transition. This transition was reported previously for the square-lattice version of CPM~\cite{Durand2019}, where it was studied as a function of temperature. Here, we show the transition as a function of the surface tension and show that it is not limited to the classical square-lattice version of CPM. By the Mermin-Wagner theorem, there should be a narrow, intermediate band of hexatic phase in the phase diagram, as also reported in the literature\cite{Durand2019, Pasupalak2020, Li2021}. However, this phase is notoriously difficult to identify and we did not carry out the detailed analysis required to establish its presence. Our focus is instead on the effect of lattice artifacts on the overall trends.

Bearing this caveat in mind, our transitions have the hallmarks of a first-order phase transition. These include a sharp change in the value of the hexatic order and the diffusion coefficient, and the presence of what appears to be coexistence between ordered and disordered cell populations, as shown in Fig.~S1.
One should be aware that the simulations are in $NVT$ ensemble, thus, the presence of a coexistence region tends to smoothen averaged curves. To establish the coexistence behavior proper free-energy calculations would have to be performed, but this was not the main goal of our present work.

The value of the isoperimetric quotient at the transition, see Fig.~S5, lies between that of a regular pentagon and a regular hexagon, which is commensurate with this observation. Additionally, as is plotted in Fig.~\ref{fgr:Phase_transition_hex}a, the transition value of hexatic order matches with the value reported in Ref.~\citenum{Durand2019}, providing additional support for the similarity between our findings and theirs. Other studies in the literature that used `deformable' particles,~\textit{e.g.}, phase-field~\cite{Monfared2023, loewe2020} and Fourier contour~\cite{saito2023active} approaches, have also reported the emergence of hexagonal solid phases in model tissues, lending further credence to our result.

This makes the transition in the CPM markedly different from the jamming transitions reported for the vertex~\cite{Thomas2023, Bi2015} and active Voronoi models~\cite{Atia2018, Bi2016}. 
These appear to leave the structure amorphous and occur at an average (target) value of $q = q_{\mathrm{reg}}(5)$. The regular (perfect) pentagon is a shape that cannot tile the plane confluently. Thus, imposing this on all cells in the model tissue leads to frustration between local and global constraints, resulting in a disordered arrest. Considering the similarities between cellular-Potts, phase-field, and Fourier-contour models, we surmise that a lack of border flexibility may be partly responsible for the difference. That is, in the vertex and Voronoi descriptions, the cells are described as polygons whose dynamics is ruled by their vertices and centroids, respectively. This means that they have far fewer degrees of freedom to their dynamics when compared 
to CPM at an equal cell number.

In this context, the work by Sadhukhan and Nandi~\cite{Sadhukhan2021} should be mentioned, who have reported glassy behavior in the CPM with a Hamiltonian restricting the perimeter of the cells, as well as their area. This choice brings their version of CPM closer to the vertex and Voronoi models. However, we believe that the observation of square cells by these authors may be attributed to lattice artifacts~\cite{Mare}. For their large preferred perimeter regime, the \textit{interlocking} of cells can also be attributed to the underlying lattice. As a lattice-based model, CPM has the inherent weakness of dealing with target perimeter lengths.
Our work most closely follows that of Chiang and Marenduzzo~\cite{Chiang2016} who identify a glass transition for the square-lattice CPM. However, in reproducing their results for the case of no self-propulsion, we have conclusively shown that this glassy behavior is not present. We did this by studying the hexatic orientational order parameter across the transition. Lastly, we turn to the results of Ref.~\citenum{Devanny2023}. They report a similar value of the shape parameter at the transition. However, they identify this transition as a jamming one. This is likely a misinterpretation of their findings, and this issue could be resolved by examining the hexatic order in their `jammed' phase.

\subsection{\label{sub:experiment}Connection to Experiment}

There have been many studies focused on jamming transitions and glassy behavior in living tissues,~\textit{e.g.}, reviewed by Refs.~\citenum{Blauth2021, LawsonKeister2021}. However, these are not the only arrested patterns that can exist in these systems. Hexagonally packed cell arrangements are also frequently found in epithelia across a range of species and developmental stages~\cite{Cislo2023, Sun2019, Classen2005, Pilot2005, Aigouy2010, Hayashi2004, Togashi2024, Lecuit2007, Gibson2006, Ortolan2022, Napoli2021, Islam2023, Jain2020, Lemke2021, Zallen2004, Hocevar2009, Johnson2021}. Although more complicated mechanisms such as adherens junctions~\cite{campas2024} can play a role in maintaining cellular order in hexagonal cell packings, it is informative to distinguish models that can capture this type of packing by their nature. 
The importance of this study is that we showed that regular and irregular versions of CPM, show an order/disorder transition, above which they were able to readily achieve the crystallized configuration. This makes them suitable for studying tissues with hexagonal cell packing.

\subsection{\label{sub:artifacts}Resolving lattice artifacts}

The main advantage we derive from our irregular underlying lattice is to rid our simulations of two types of (obvious) lattice artifacts, without having to resort to higher-order neighbor coupling, as has been discussed in,~\textit{e.g.}, Ref.~\citenum{Mare}. Figures~\ref{fgr:Phase_transition_hex} and~S3 both clearly provide evidence that a square lattice CPM with a Moore neighborhood rule can lead to an abnormal distribution of hexatic order at high surface tension. This happens because on the square lattice, the cells are essentially forced to assume the rhombus-like shape that is shown in Fig.~\ref{fgr:Phase_transition_hex}b, snapshot 7. There are lattice artifacts on the hexagonal lattice too, although they appear not to be as severe as they are on the square lattice. 
The cells become hexagonal, but these hexagonal cells are forced to align with the underlying lattice. The disordered lattice is free of both of these artifacts within the error. Firstly, the distribution of hexatic order for the simulations on the disordered lattice is completely smooth and free of any unphysical peak within error. Secondly, there is no obvious preferred direction for cell borders on the irregular lattice.  

By resolving these two lattice artifacts, we decreased the influence of the lattice geometrical symmetries on the shape of cells. We should mention that artifacts are an inevitable feature of lattice-based models. The shape of cells is always to some extent dependent on the geometry of the underlying lattice. However, as we have shown, by breaking the orientational symmetries of the underlying lattice, one can significantly weaken this dependency.
This is an important advantage because as discussed in several studies~\cite{Park2015, Luciano2022, gottheil2023, alert2021, Atia2018}, the shape of the cells can be strongly connected to the dynamics of the tissue.
\subsection{\label{sub:efficiency}Efficiency and Considerations}

Our irregular-lattice version of CPM not only is free of demonstrable artifacts but it can also be applied to any lattice that derives from a Voronoi construction, including regular ones. This provides our description with greater flexibility. In fact, for simulations on the hexagonal lattice, we used the same code, but with a perfect hexagonal Voronoi lattice as an input. When using an irregular lattice, it may be prudent to generate several independent realizations of the lattice. This should further reduce the correlations when taking statistical averages; note that we used only one disordered lattice in the current study.

In terms of computational efficiency, an extra run time is incurred using our approach compared to the square-lattice CPM. This was, however, very reasonable, as the average simulation time per $10^6$ MC sweeps for the square-lattice CPM was generally in the range of 7 to 11 hours, while it was in the range of 8 to 15 hours for our generalized CPM. This timing data was obtained using cluster nodes equipped with, on average, 20 CPU cores and 64 GB of RAM. Even in the worst case, the run time was less than twice that of the regular CPM.

Note that to properly evaluate the perimeter of our cells, we used Voronoi tessellation on their centers of mass. Even on the irregular lattice, the contour length of the cells is artificially high, as we established by examining a very large circular cell in isolation. This is a limitation of the model, which could be overcome by using alternative perimeter evaluation algorithms such as Voronoi tessellation.

\subsection{\label{sub:shapes}Shape-Parameter-Based Characterization}

Several studies have looked at distributions of properties of the cell neighborhood. Some considered the aspect ratio of the cells~\cite{Atia2018, Sadhukhan2022}, while others extracted the number of neighbors~\cite{Farhadifar2007, Roshal2023, SnchezGutirrez2015, Zallen2004}. Here, we calculated the distribution underlying the mean isoperimetric quotient, and from these, we extracted fractions of pseudo-$n^{\ast}$-gons to gain a deeper insight into the system. Figure~\ref{fgr:N_star_fractions_a_b} for the irregular and Fig.~S4 for the regular lattices, respectively, show that the majority of the polygons in the subdiffusive state are pseudo-hexagons, while in the sufficiently fluid-like regime, they are pseudo-pentagons. Close to the transition, we have observed that their relative abundance in the sample crosses over.

We found a single study by Saito and Ishihara~\cite{saito2023active}, who examined isoperimetric-quotient distributions across the transition in their Fourier-contour model. They also reported a unimodal-to-bimodal transition in the distributions, as we have observed. The bimodality in the fluid-like regime is indicative of a coexistence between more-rounded and less-rounded cells, which is understood to lead to fluidity of the tissue. We note that in the current literature~\cite{Chiang2016,loewe2020, Sadhukhan2021, Devanny2023, saito2023active, Zhang2020} there is a tendency to focus on the mean value of the shape parameter as the indicator of transition. This is undoubtedly motivated by its relevance to the behavior of the vertex and Voronoi models~\cite{Bi2016, Damavandi2022, Krajnc2020, Barton2017, Yan2019, Huang2022, Bi2015, Li2018, Giavazzi2018, Atia2018, Yang2017, Thomas2023}. However, as we have argued above, these models fall into a different class. We, therefore, believe it to be informative in general to examine shape-parameter distributions rather than only means.

As a complementary quantity to describe the shape of the cells we studied the circularity parameter $C$~\cite{uni2008, uni2010}, which is in essence, one of Hu moment invariants~\cite{MingKueiHu1962}. This compares the area of shapes to their moment of inertia, rather than to their perimeter, which is the case for $q$. This makes $C$ less sensitive to noise at the boundary or definition issues with perimeter length. That is, we could have performed the circularity analysis on the original cell shapes rather than their Voronoi-based counterparts. However, we found that applying Voronoi tessellation to the cells led to a better comparison between the two roundness measures. 

\section{\label{sec:summary}Summary and Outlook}

In this study, we extended the regular (square-lattice) cellular Potts model to work on arbitrary (ir)regular lattices. The main motivation of this study was to eliminate lattice artifacts observed for square-lattice CPMs in the literature. We demonstrated that such artifacts could indeed be eliminated by using an irregular lattice generated from a fluid-like state using Voronoi tessellation. Our generalized CPM maintains many of the desirable features of a base CPM, at the price of a small computational overhead.

We gained the following insights using our algorithm on both irregular, square, and hexagonal lattices. First, there is an order-disorder transition in confluent cell monolayers when using CPM on all lattices that we considered. This sets the base CPM apart from the active Voronoi and vertex models, 
in which there is a rigidity transition that maintains the disorder of the fluid-like state for shape parameters close to those of a regular pentagon. 
This makes the CPM suited to describe epithelia in which ordered cell arrangements form, as these can be reached directly and straightforwardly from the disordered state. 
This, however, does not mean to imply that the CPM can never describe glassy dynamics. For example, glassy dynamics could be realized in the CPM by modifying the Hamiltonian or by introducing restrictions on rearrangements leading to crystalline order.
Second, for the square lattice, the cell shape can be impacted by artifacts above the transition, while for the hexagonal lattice, the borders of cells are affected by the symmetry of the lattice. Our irregular-lattice CPM did not exhibit these undesirable features.

In addition, we gained deeper insight into the transition by studying the distributions of different quantities across the transition. These included the hexatic bond order, the isoperimetric quotient, and the circularity, which overlap in their descriptiveness of local neighborhoods. For the isoperimetric quotient, the transition is closely correlated to the crossover in the fraction of pentagonal and hexagonal neighborhoods found in our simulations. However, the transition does not lie at an average isoperimetric quotient of a pentagon, as is the case for the vertex and active Voronoi models. This is because, unlike these two systems, the shape is not a target parameter in the Hamiltonian describing the system. It is further important to realize that examining the mean isoperimetric quotient, or the related shape index, may give an incomplete picture of the behavior of the system. 
We exclude the athermal Voronoi model in this comparison, where the rigidity transition is known to be absent\cite{sussman_no_2018}.

Looking forward, we note that artifacts can be straightforwardly eliminated through the introduction of an irregular grid. This may have additional advantages when coupling the dynamics of the cells to that of external fields, such as food or chemical fields in bacterial colonies~\cite{Hallatschek2007}. We will explore these directions in modeling real biological tissues using our generalized CPM in the future and hope this will lead to wider adoption of our approach.

\section*{Data Availability}

Open data package containing the means to reproduce the results of the simulations available at: [https://doi.org/10.24416/UU01-IBVNT1]

\section*{Author Contributions}

The authors contributed as follows to the study: Conceptualization (JdG \& HN), Data Curation (HN), Formal Analysis (HN), Funding Acquisition (JdG), Investigation (HN), Methodology (JdG \& HN), Project Administration (JdG), Resources (JdG), Software (HN), Supervision (JdG), Validation (HN \& JdG), Visualization (HN), Writing -- Original Draft (HN), Writing -- Review \& Editing (JdG).

\section*{Conflicts of interest}

The authors declare that there are no conflicts of interest.

\section*{Acknowledgements}

The authors acknowledge financial support from the Netherlands Organization for Scientific Research (NWO) through Start-Up Grant 740.018.013. We are grateful to Bryan Verhoef, Kim William Torre, Meike Bos, and Davide Marenduzzo for useful discussions.

\bibliography{references}

\clearpage

\begin{widetext}

\section*{Supplementary Information}

\renewcommand{\thefigure}{S\arabic{figure}}
\setcounter{figure}{0}  

\subsection{Coexistence}

An example snapshot in Fig.~\ref{fgr:Coexistence} shows patterns of coexistence for a simulation with $\alpha=2$, which is close to the transition point on the disordered lattice. Figure~\ref{fgr:Coexistence}a reveals that within one simulation volume, there are regions with more disordered (yellow cells are abundant) and ordered (purple cells in the bottom-right quadrant) configurations. We have plotted Voronoi tessellated version of cells in Fig.~\ref{fgr:Coexistence}b, for which the colors are based on the hexatic order threshold $\vert \psi_{6} \vert  = 0.76$; above and below indicated using purple and orange, respectively. This representation better shows the ordered cluster.

\begin{figure}[!htb]
\centering
\includegraphics[width=\textwidth,trim=1 1 1 1,clip]{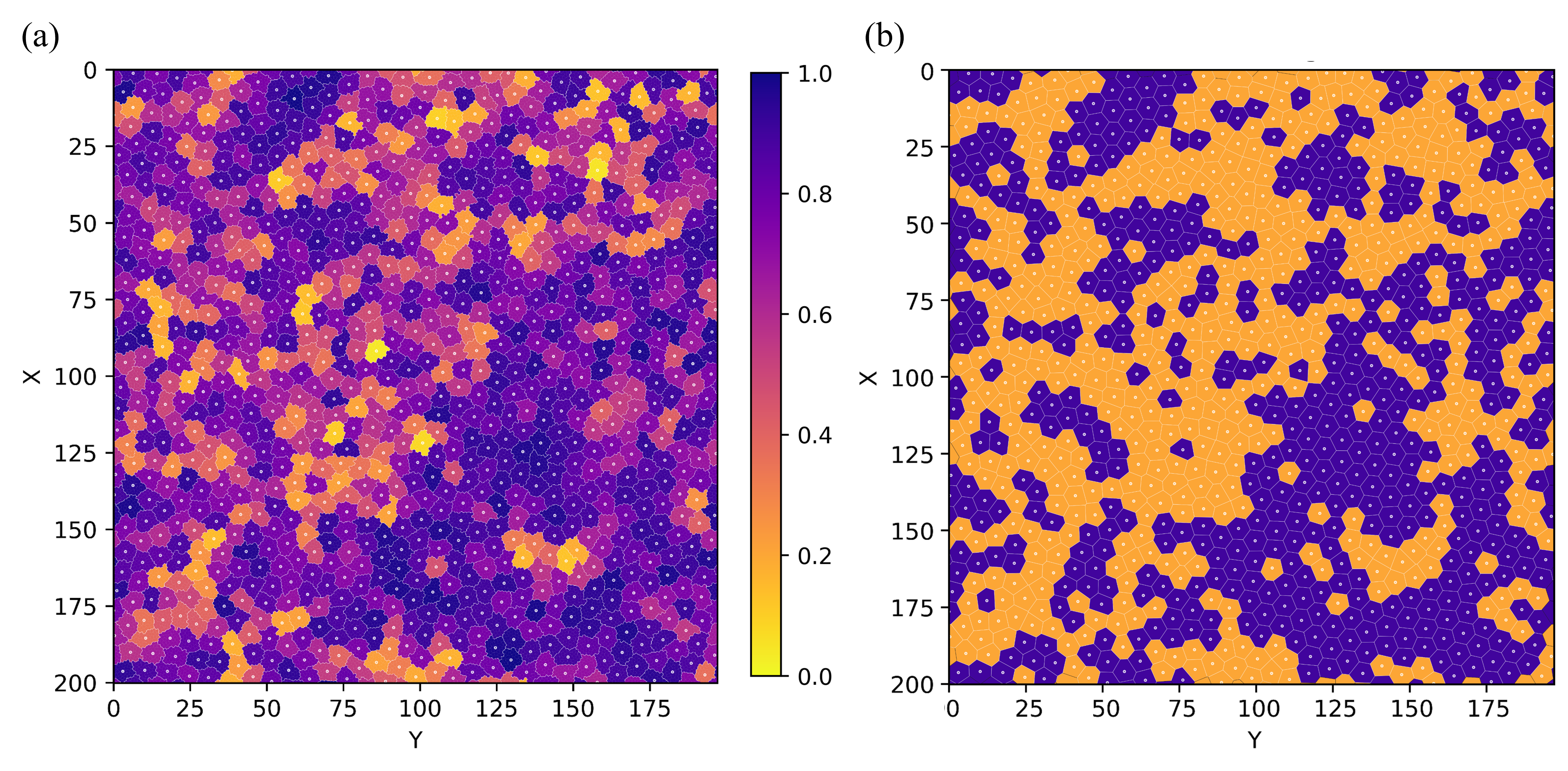}
\caption{\label{fgr:Coexistence}An example simulation snapshot revealing that an ordered and disordered phase are present in the same simulation volume for $\alpha = 2.0$. 
(a) Cells are colored by their local hexatic order $\vert \psi_6 \vert$ bar, as indicated by the bar on the right-hand side of this panel.
(b) The Voronoi-tessellated version of the same snapshot in panel (a). In this panel, the purple and orange cells have a hexatic order greater and smaller than $0.76$, respectively. The colors in panel (b) are not related to the color bar used in (a). 
The white dots in both panels show the center of mass of the cells.}
\end{figure}

\newpage
\subsection{Mean square displacement (MSD)}

Figure~\ref{fgr:MSD_plot_irr} shows the mean-squared displacement (MSD) of the center-of-mass of cells on the disordered lattice for all values of $\alpha$ that were studied. They are averaged over the individual cells across the various independent simulations that were performed at a given state point. 
We can see that for low values of $\alpha$, ($\alpha<2.0$), the MSD is diffusive for the times shown here. For higher values of $\alpha$, there is an initial plateau, which is due to caging effects. The MSD eventually becomes diffusive, though for some systems it can take a very long time. We only simulated trajectories close enough to the cross-over region between ordered and disordered for longer than $10^{6}$~MCS.

\begin{figure*}[!htb]
\centering
\includegraphics[width=0.8\columnwidth,trim=1 1 1 1,clip]{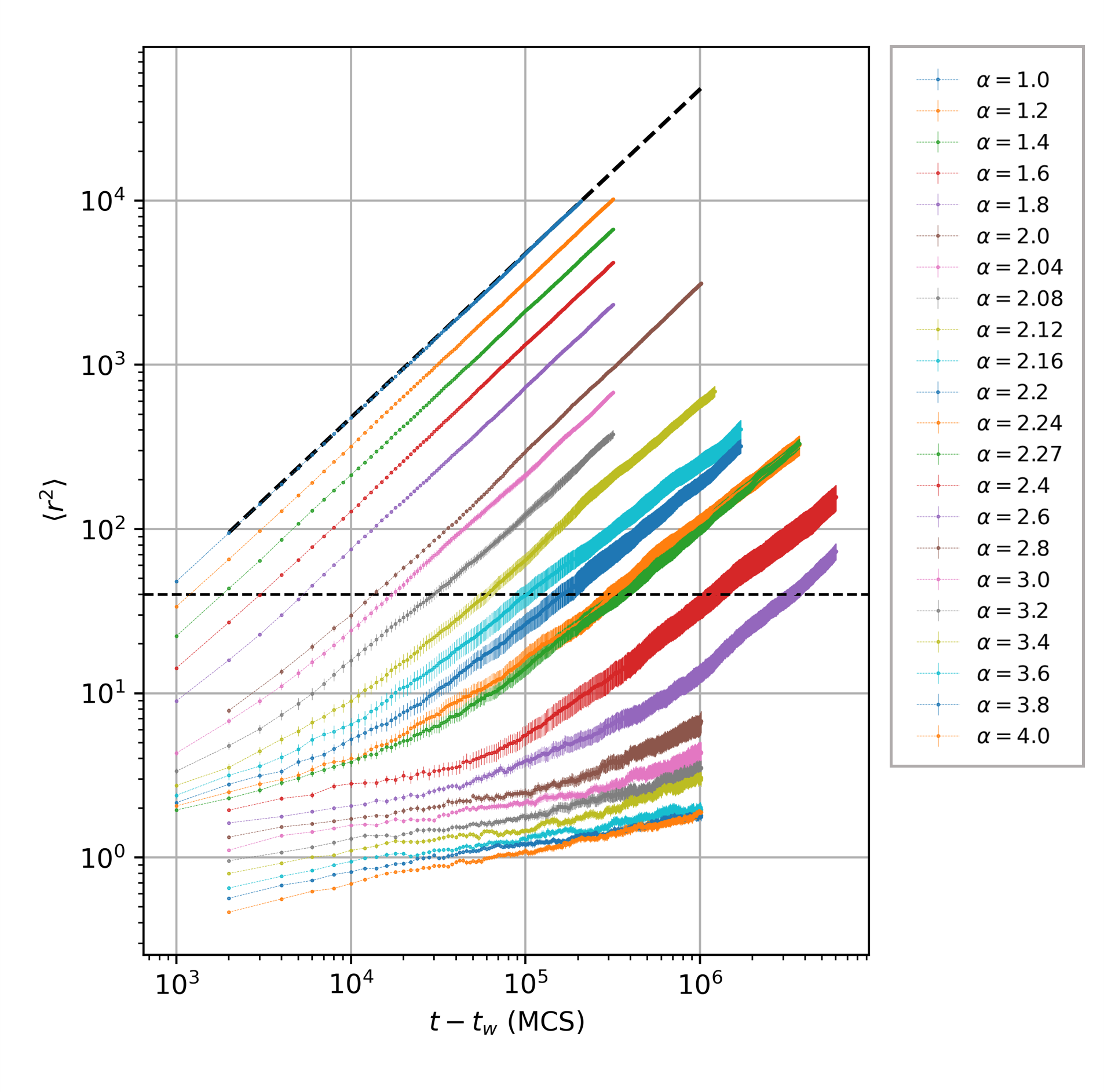}
\caption{\label{fgr:MSD_plot_irr}The average mean-squared displacements $\langle r^{2} \rangle$ of the center-of-mass of cells are plotted as a function of simulation time $t$ (starting from the waiting time $t_{w}$) for different values of $\alpha$. All simulations were performed on a disordered lattice. The legend provides the values of $\alpha$, but in reading the graph $\alpha$ increases from top to bottom. Measured points are connected to guide the eye and the error bars indicate the standard error of the mean. The horizontal dashed line indicates $A_0$, the target cell area. This can be used to determine the time at which a cell has displaced its own size on average. The sloped dashed line indicates a diffusive trend and partially overlaps with the $\alpha = 1.0$ data, which is thus clearly diffusive.}
\end{figure*}

\newpage

\subsection{Lattice artifacts}

Figure~\ref{fgr:sq_artifacts_suppl} shows the probability density function (PDF) belonging to the hexatic order parameter $\vert \psi_{6} \vert$ as measured for a high value of $\alpha = 4$. The distribution shows a clear main peak lattice artifact for the underlying square lattice, with a potential secondary artifact distribution to the right of the vertical black line. Conversely, the distribution on the irregular lattice is smooth and shows no clear lattice artifacts.

\begin{figure}[!htb]
\centering
\includegraphics[width=0.5\columnwidth,trim=6 1 1 1,clip]{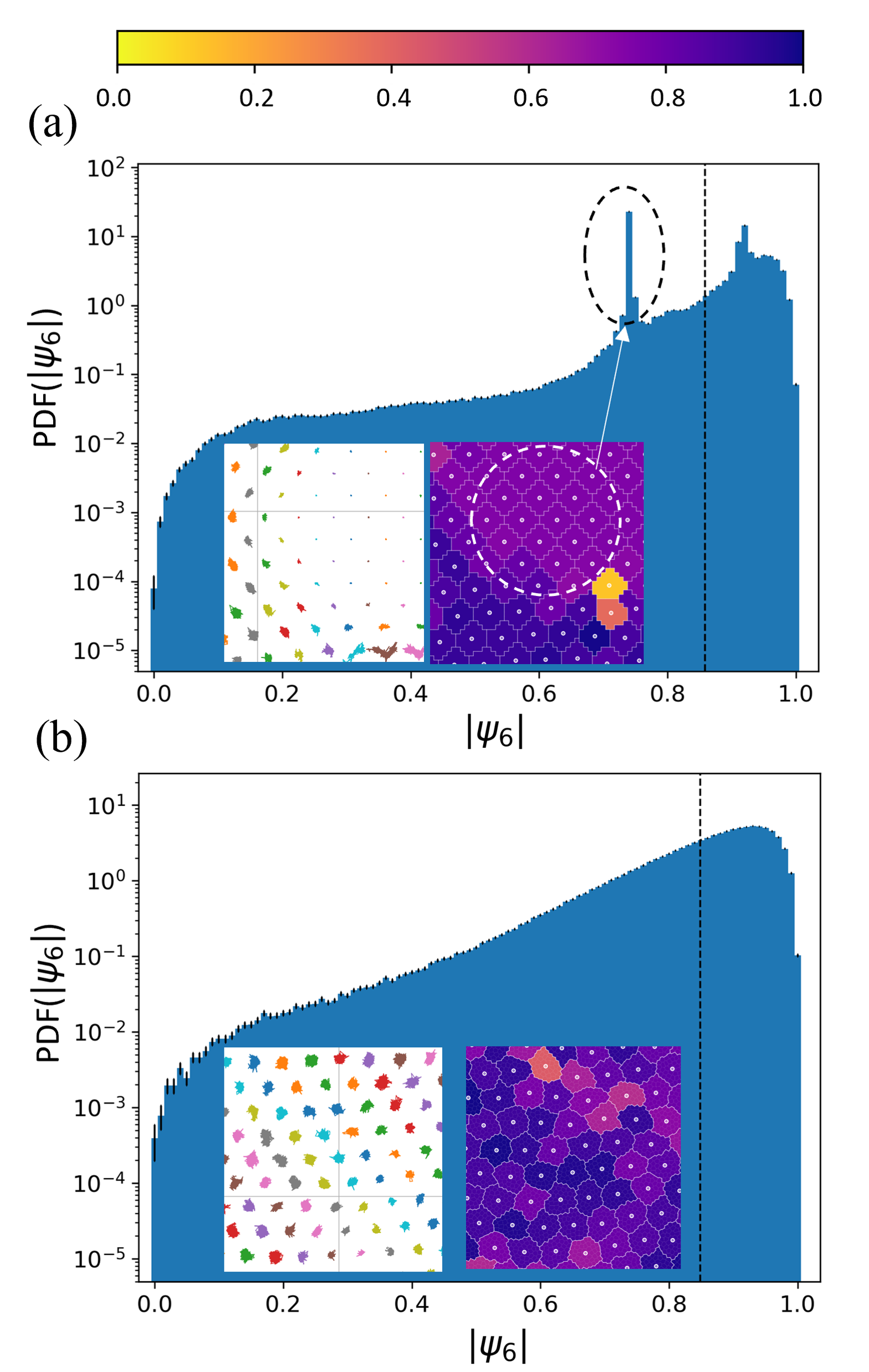}
\caption{\label{fgr:sq_artifacts_suppl}The lattice artifacts observed on the square lattice disappear on the disordered lattice. (a) The distribution (probability density function; PDF) of the hexatic order on the square lattice for $\alpha=4$. The unphysical peak is indicated by the dashed ellipse. (b) The same distribution for the same value of $\alpha$ on the disordered lattice. The insets in both panels show the configurations of the cells and the trajectories of their centers of mass. On the right-hand side of the inset, the model cells are colored by the hexatic order parameter $\vert \psi_{6} \vert$, for which the color bar is shown at the top of the figure. The coloring on the left-hand side is to distinguish individual cells and has no physical significance. In both graphs, the error bars indicate the standard error of the mean and the vertical dashed line represents the average value of the distribution.}
\end{figure}

\clearpage

\subsection{Pseudo-polygons on regular lattices}

Figure~\ref{fgr:crossover_sq_hex} shows the fractions of pseudo-polygons on square and hexagonal lattices. From Fig.~\ref{fgr:crossover_sq_hex}a, it becomes clear that the argument we made about the crossover between $f_6$ and $f_5$ being an indicator of the phase transition on the irregular lattice, holds also for the square lattice. Figure~\ref{fgr:crossover_sq_hex}b shows the fractions on the hexagonal lattice.  As can be seen, on the square lattice, as with the irregular one, the transition region well matches with the crossover point between the fractions of pseudo-hexagons and pseudo-pentagons. However, for the hexagonal lattice, there is a slight mismatch. Though the mismatch is small, this could suggest that there is an anomalous diffusion behavior, caused by the underlying hexagonal order. That is, the system becomes subdiffusive, before crystallization sets in.

\begin{figure}[!htb]
\centering
\includegraphics[width=0.5\columnwidth,trim=1 1 1 5,clip]{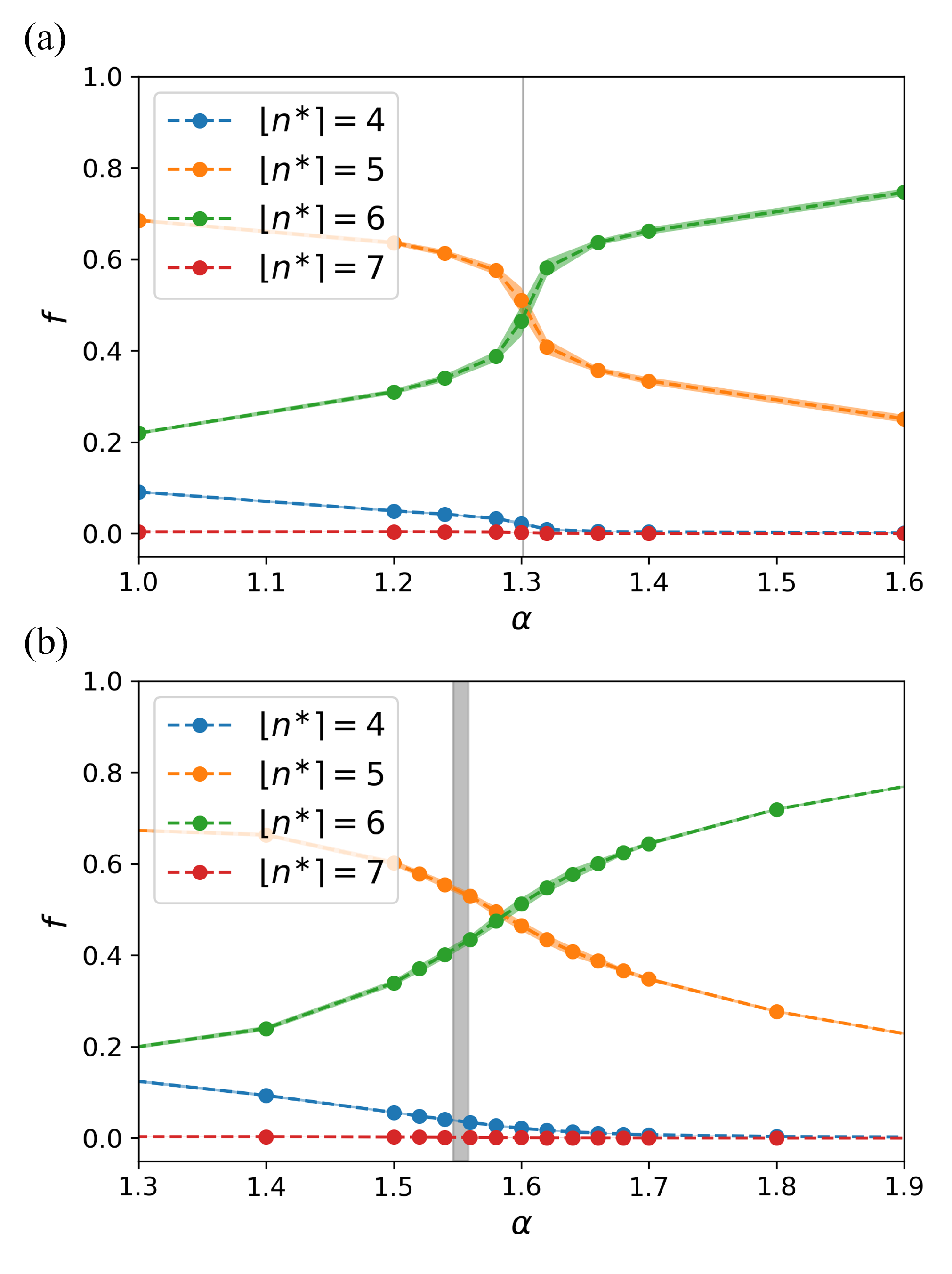}
\caption{\label{fgr:crossover_sq_hex}
The fractions of pseudo-polygons $f_{n^{\ast}}$ as a function of the surface tension $\alpha$ obtained for square (a) and hexagonal (b) lattices. The number of edges for the pseudo-polygons is indicated in the legend and the colored area around the respective curves indicates the standard error of the mean. The vertical gray line (bar) shows  where the diffusion coefficient drops, also see many text.}
\end{figure}

\clearpage

\subsection{$q_{V}$ at the transition} 
The fractions shown in Fig.~7a in the main text derive from integrals of the $q_{V}$ distributions. We can also examine the average and the median value of $q_{V}$ across the transition zone. In 
figure~\ref{fgr:beta_vs_qV_mean_median}a, 
we find that for the irregular and square lattices, the median of $q_{\rm{V}}$ being equal to $q(n^{\ast} = 5.5)$ is a very good indicator of the transition,~\textit{i.e.}, to within the standard error of the mean. However, for the hexagonal lattice, there is a mismatch, which is likely similar in origin to the one observed in Fig.~\ref{fgr:crossover_sq_hex}b. Figure~\ref{fgr:beta_vs_qV_mean_median}b provides the average ($\langle q_{\rm{V}} \rangle$). Here, we clearly see that neither proposed structural measure provides an accurate location for the transition. Thus, we conclude that the median of $q$ is the more relevant quantity.

\begin{figure}[!htb]
\centering
\includegraphics[width=0.5\columnwidth,trim=3 1 1 5,clip]{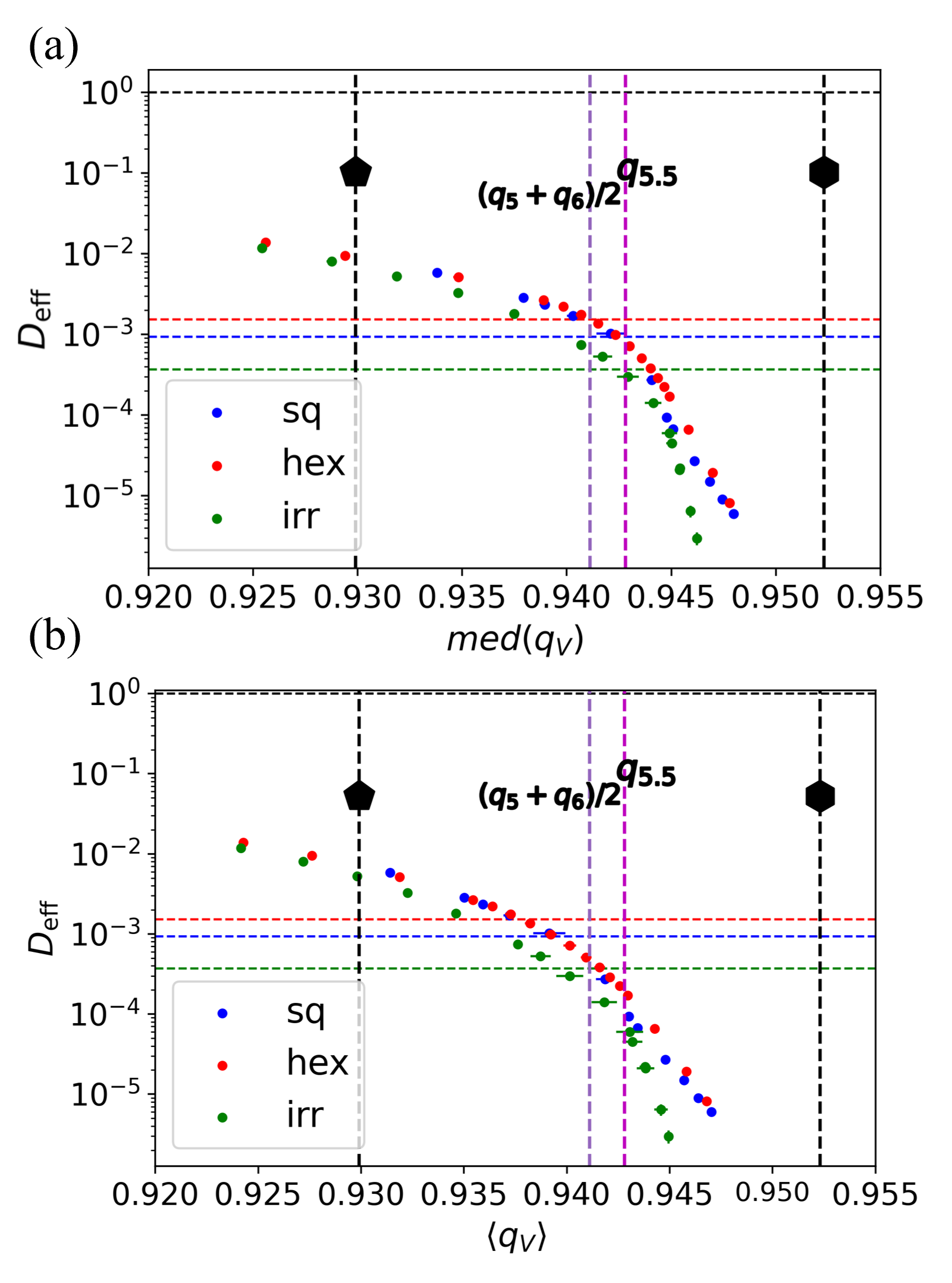}
\caption{\label{fgr:beta_vs_qV_mean_median}Behavior of the median and average of $q_{V}$ across the transition.
(a) The diffusion coefficient is plotted as a function of the median of $q_{V}$. (b) The diffusion coefficient as a function of the average of $q_{V}$. The three underlying lattices are as indicated using the colors in the legends and the error bars indicate the standard error of the mean. The values of $q_{V}$ for a pentagon and hexagon are indicated using vertical dashed lines and symbols. The derived values for a generalized ($n^{\ast} = 5.5$)-gon and the arithmetic mean of the hexagon and pentagon values are indicated using vertical dashed magenta and purple lines, respectively.  The horizontal dashed lines indicate $D^{\blacksquare}_{\rm{eff}}$ on different lattices.}
\end{figure}

\clearpage

\subsection{$n^{\ast}$ on regular lattices}

Figure~\ref{fgr:qV_extrema_sq_hex} shows the distribution of $n^{\ast}$ parameter and the behavior of its extrema on the square and the hexagonal lattices. Like the irregular lattice, we see that the number of the extrema of this distribution determines the state. As we discussed for the irregular lattice, the local minimum and the minor local maximum merge together and disappear at the transition, while the position of the global maximum experiences an abrupt change in its slope, and hardly changes in value beyond the transition. The global maximum at the transition is around $n^{\ast} \approx 5.7$, as was the case for the irregular lattice.

\begin{figure*}[!h]
\centering
\includegraphics[width=\textwidth,trim=8 3 3 5,clip]{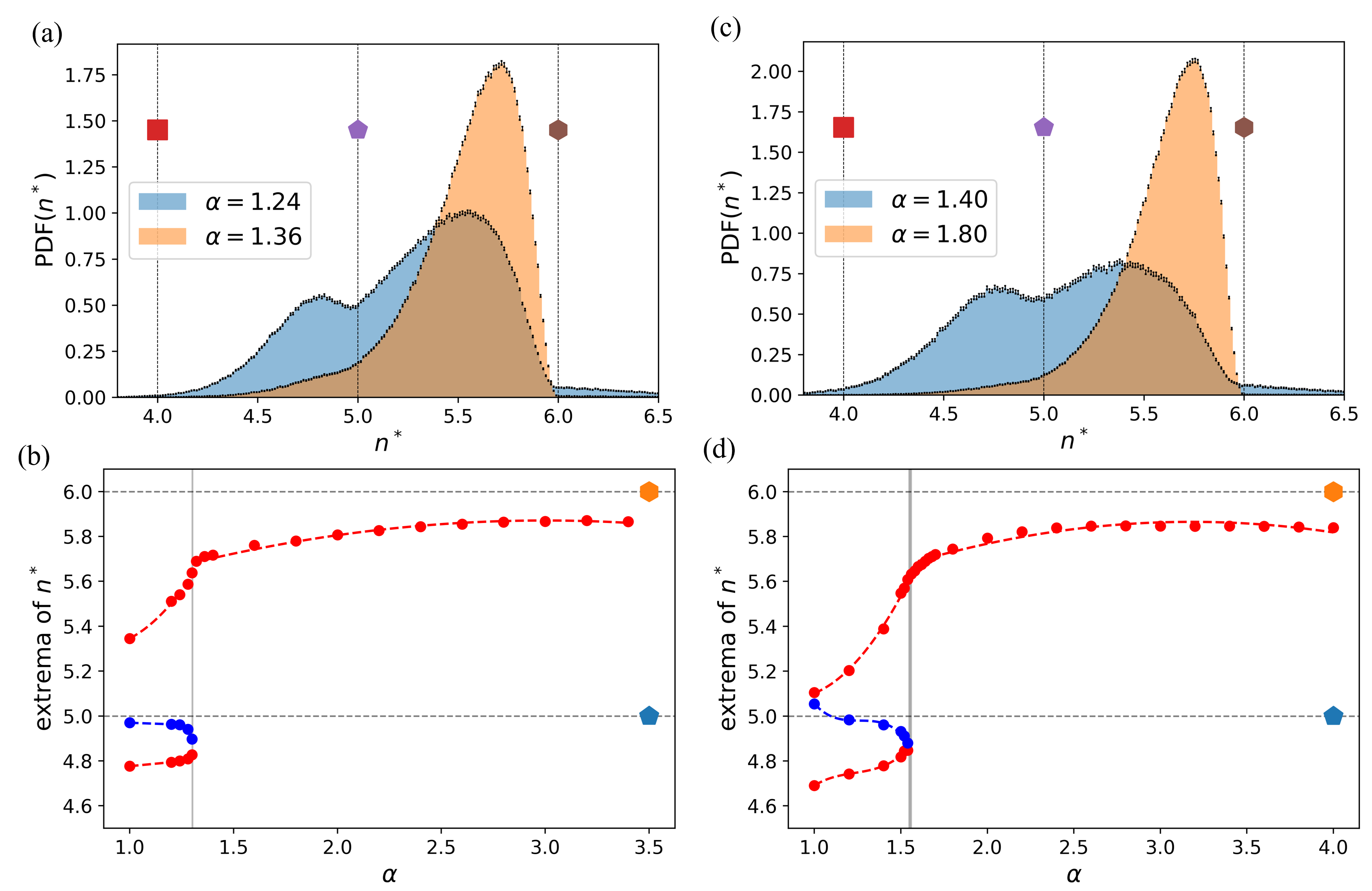}
\caption{\label{fgr:qV_extrema_sq_hex}
The distribution of $n^{\ast}$ parameter indicates the phase transition on regular lattices, but less so on hexagonal lattices. (a) Distribution of $n^{\ast}$ for two values of $\alpha$ are shown as well as the local minimum on the square lattice. For $\alpha=1.24$ (blue) the tissue is fluid-like, while for $\alpha=1.36$ (orange) it is solid-like. (b) The positions of the maxima (red) and the local minimum (blue) of $n^{\ast}$ distribution on the square lattice are plotted as a function of $\alpha$. The dashed lines are there to guide the eye, and the gray box indicates the transition point. (c,d) The analog of (a,b) for a hexagonal lattice, with a typical fluid-like ($\alpha=1.4$, blue) and solid-like state ($\alpha=1.8$, orange).}     
\end{figure*}

\clearpage

\subsection{Circularity on regular lattices}

Figure~\ref{fgr:cV_mode_sq_hex} shows the behavior of the mode of the Voronoi circularity on square and hexagonal lattices. Unlike the distributions of $q_V$ and $n^{\ast}$, the distribution of $C_V$ always has only a single peak. Nonetheless, the mode of this distribution is also a complementary determinant of the transition. As for the irregular lattice we discussed in the main text, on the square and hexagonal lattices, the mode of $C_V$ distribution is very close to the circularity of the pentagon, when the system transitions from disordered and fluid-like to ordered and solid-like.

\begin{figure*}[!h]
\centering
\includegraphics[width=\textwidth,trim=10 3 3 5,clip]{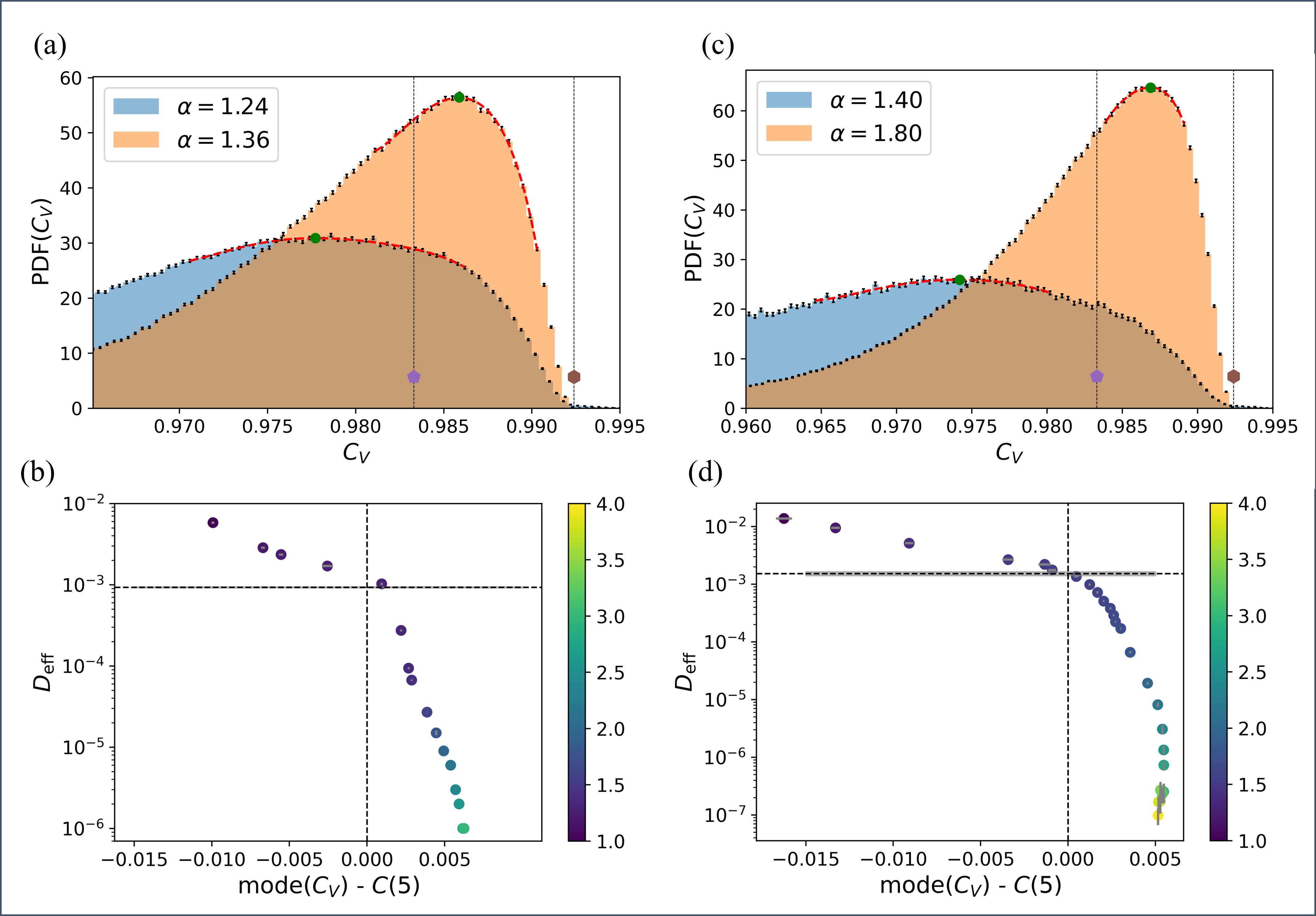}
\caption{\label{fgr:cV_mode_sq_hex} The mode of the Voronoi-based circularity indicates phase transition on regular lattices, (a) The PDF of the Voronoi-based circularity, $C_V$, for two values of $\alpha$, one of which in fluid-like ($\alpha=1.24$, blue) and the other in solid-like ($\alpha=1.36$, orange) regime, on the square lattice. The positions of the peaks (green dots) are determined by polynomial fittings (red dashed lines). The vertical dashed lines show the values of circularity for the pentagon and hexagon. (b) The diffusion coefficient, $D_{\rm{eff}}$, plotted as a function of the departure between the mode of $C_{V}$ from the circularity of a pentagon, $C(5)$, on a square lattice. 
The horizontal dashed line shows the diffusion coefficient at transition,~\textit{i.e.}, $D^{\blacksquare}_{\rm{eff}}$.
The data points are colored by the values of $\alpha$ as shown by the color bar on the right. The data of the highest 3 values of $\alpha$ are not plotted because of the considerable lattice artifacts. (c,d) The analog of (a,b) for a hexagonal lattice, with a typical fluid-like ($\alpha=1.4$, blue) and solid-like state ($\alpha=1.8$, orange). 
The gray box in panel d indicates the standard error of $D^{\blacksquare}_{\rm{eff}}$. This plotted in panel b as well, but the error is too narrow to properly visualize.}
\end{figure*}

\newpage

\end{widetext}

\end{document}